\documentclass[onecolumn,preprintnumbers,amsmath,amssymb]{revtex4}

\usepackage{graphicx}% Include figure files
\usepackage{color}
\usepackage{bm}% bold math

\newcommand{\ve}[1][K]{\mathbf{#1}}

\begin{document}

\title{Non Markovian polymer reaction kinetics}

\author{T. Gu\'erin}
\affiliation{Laboratoire de Physique Th\'eorique de la Mati\`ere Condens\'ee, CNRS/UPMC, 
 4 Place Jussieu, 75005 Paris, France.}

\author{O. B\'enichou}
\affiliation{Laboratoire de Physique Th\'eorique de la Mati\`ere Condens\'ee, CNRS/UPMC, 
 4 Place Jussieu, 75005 Paris, France.}

\author{R. Voituriez}
\affiliation{Laboratoire de Physique Th\'eorique de la Mati\`ere Condens\'ee, CNRS/UPMC, 
 4 Place Jussieu, 75005 Paris, France.}

\date{02/07/2012}
%\nodate

\begin{abstract}
Among transport--limited reactions, reactions involving  polymeric chains play an important role. Both intramolecular reactions such as cyclization and intermolecular reactions have been extensively studied experimentally and theoretically, and have been shown to lead to complex kinetics. Despite  these considerable efforts, there is to date no exact explicit analytical treatment of transport--limited polymer reaction kinetics, even in the case of the simplest model of flexible polymer - a phantom Rouse chain of monomers connected by linear springs. The main difficulty arises from the fact the  motion of a single monomer in the chain  is non Markovian. Here, we introduce a new analytical approach to calculate the mean  reaction time of polymer reactions that  encompasses the non Markovian dynamics of the problem.  A key step of our method  relies on the determination of the statistics of the polymer conformation at the very instant of reaction, which provides as a by product  new information on the reaction path. We show that the typical reactive conformation of the polymer is  more extended than the equilibrium conformation, which leads to reaction times significantly shorter than predicted by existing Markovian theories. Together, these results provide a better understanding of the complex kinetics of polymer reactions  involved for example in the formation of loops of RNA or polypeptides chains.
\end{abstract}

\maketitle

Reactions involving macromolecules and in particular polymer chains are ubiquitous.   An important example of  intramolecular polymer reaction is provided by  cyclization reactions, which consist in forming a loop joining the two ends of a polymer. Such reactions have been widely studied  both theoretically \cite{WILEMSKI1974a,WILEMSKI1974b,Szabo1980,Friedman1989,FRIEDMAN1993b,FRIEDMAN1993,DEGENNES1982,Toan2008} and experimentally  \cite{Bonnet1998,Lapidus2000,Wallace2001,Kim2006,Wang2004,Moglich2006,Buscaglia2006,Uzawa2009}, mainly because of their  relevance to biological processes. Indeed, the formation of loops and hairpins in DNA is a key process in the regulation of gene expression \cite{Allemand2006};
 in the context of protein folding, the cyclization of a polypeptide chain can be seen as an elementary step of  the folding pathway \cite{Lapidus2000}. Among  intermolecular polymer reactions,  search processes involving a polymer chain and a given target, be it a catalytic site or a pore in a confining cavity, play a prominent role, as exemplified by gene delivery or viral infection, which involve a step that is kinetically limited by the search for a nuclear pore by a nucleic acid \cite{Wong2007,Dinh2007,Dinh2005}. 

The theoretical description of polymer reaction kinetics requires to take into account the intrinsically complex dynamics of a polymer chain. The motion of a monomer depends on the dynamics of the entire chain, and therefore cannot be described as a Markov process. This non Markovian feature induces the emergence of multiple time scales, and can lead to subdiffusion  \cite{KhokhlovBook,DoiEdwardsBook} and  non trivial reaction kinetics \cite{DEGENNES1982,Nechaev:2000fk,OSHANIN1994}. Numerous studies have been devoted to the theoretical analysis of polymer reaction kinetics, but until now all available explicit results  rely on Markovian approximations of this non Markovian problem. 

The benchmark theory in the field has been developed by Wilemski and Fixman \cite{WILEMSKI1974b,WILEMSKI1974a}, and   assumes that all the hidden degrees of freedom of the polymer reach instantaneously their equilibrium distribution. Another classical theory is  the harmonic spring model, where the whole polymer chain is modeled by a single spring, with an effective stiffness that takes into account the entropic stiffness of the chain \cite{Szabo1980,SUNAGAWA1975,DOI1975}. 
A third theoretical approach is based on  the renormalization group theory and leads for infinitely long  chains to perturbative results  in $\varepsilon=4-d$, where $d$ is the space dimension \cite{Friedman1989,FRIEDMAN1993b,FRIEDMAN1993}.  
More recent contributions  include  a formal  iterative solution in dimension 1 \cite{Likthman2006}, or a more refined treatment of the correlations \cite{Sokolov2003}.
These theories capture some features of the anomalous polymer dynamics, but  explicit results invariably make use of a Markovian approximation.
This assumption inevitably leads to a restricted range of applicability of these approaches, and clearly fails as soon as the reaction time is of the same order as the polymer relaxation time. In fact, results of numerical simulations have proved to significantly differ from the available theoretical predictions in a broad range of parameters \cite{Pastor1996,Toan2008,Ortiz-Repiso1998,Ortiz-Repiso1998_1}.  

Here, we propose a new approach that directly deals with the non Markovian character of the problem.  
The key step of our method relies on the determination of the statistics of the polymer conformation at the very instant of reaction, which has been disregarded so far. We show that typical reactive conformations are in marked contrast with the equilibrium conformations as is implicitly assumed in existing  Markovian theories. Our analytical approach provides a very accurate determination of the mean reaction time for both intra and inter molecular reactions %(see Fig. \ref{Fig1sketch}) 
valid for any range of parameters, which  significantly improves standard polymer reaction kinetics theories.  These results open new perspectives in the understanding of  the complex kinetics of polymer reactions. % such as RNA hairpin formation 

 \begin{figure}[h!]
 \includegraphics[width=8cm,clip]{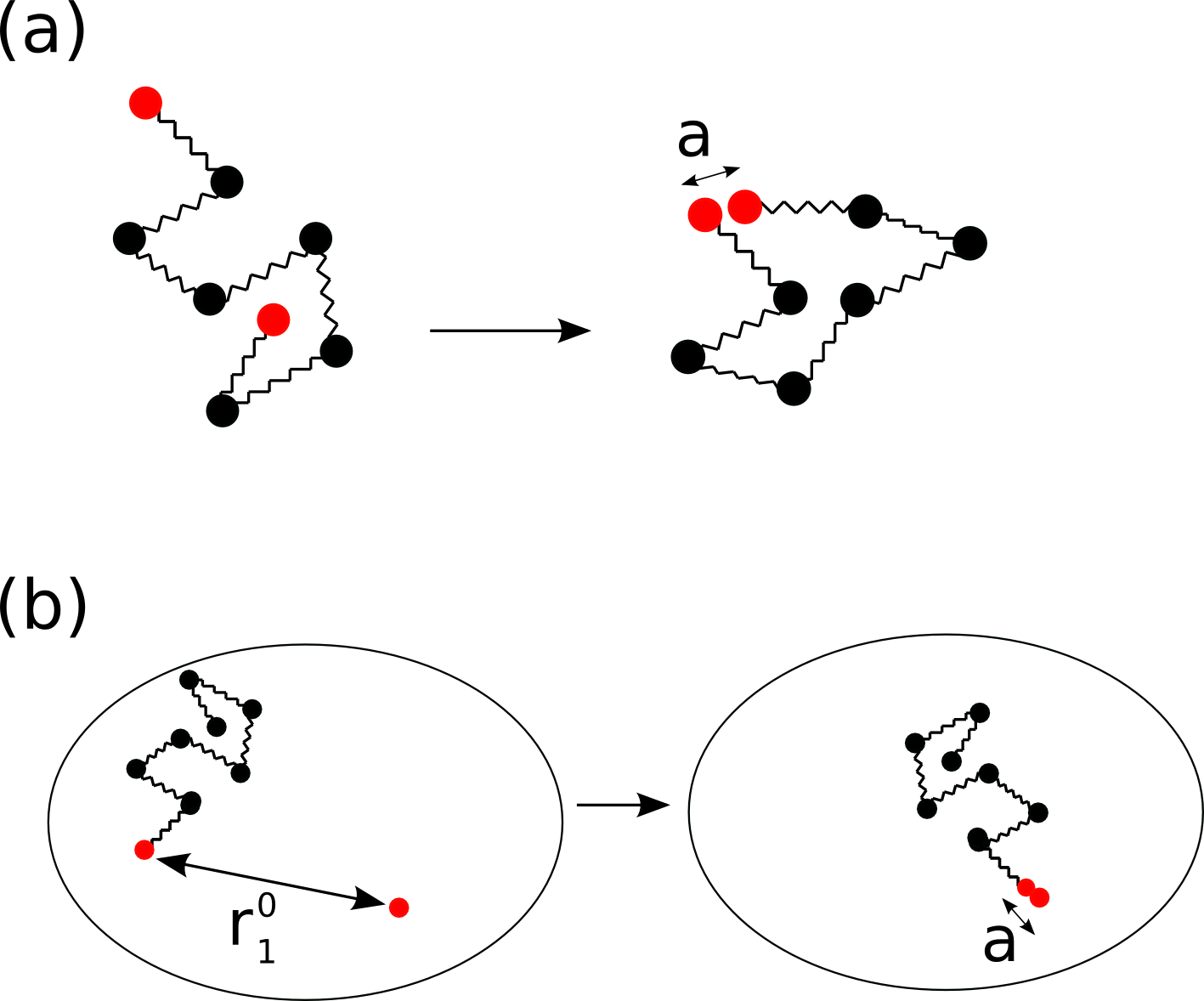}   
 \caption{{\bf Sketch of the examples of intramolecular and intermolecular reactions studied in this paper.} What is the influence of the presence of various monomers on the reaction  kinetics between reactants attached to particular monomers ? We address this question in this paper by considering the two examples of the cyclization reaction (a) % in a Rouse chain
and the reaction between a reactant attached to the end of a polymer and a fixed target % of size $a$ 
in a confining volume (b).}
\label{Fig1sketch}
\end{figure}  

%%% The theory   
\newpage
\vspace{1cm}
{\bf Results}

We consider the classical model of a  Rouse chain of $N$  monomers connected  by linear springs of stiffness $k$. The monomers experience a frictional drag of coefficient $\zeta$ and   diffuse with a diffusion coefficient $D=k_BT/\zeta$ in the force-field created by their neighbors. Even if this minimal model  neglects both hydrodynamic interactions and excluded volume effects, it captures the main features of polymer dynamics  \cite{KhokhlovBook,DoiEdwardsBook}. We denote the microscopic time scale by $\tau_0=\zeta/k$, which is the typical relaxation time of a bond in the polymer, and the microscopic length by $l_0=\sqrt{k_B T/k}$, which is the typical length of a bond. We introduce the positions $\ve[r]_i,i\in\{1,...,N\}$ of the $N$ monomers, where quantities in bold stand for vectors in the $d$--dimensional space. The probability $P(\ve[r]_1,...,\ve[r]_N,t)$ to find the polymer chain in a given configuration at time $t$ satisfies the Fokker-Planck equation  \cite{KhokhlovBook,DoiEdwardsBook}:
\begin{equation}
	\partial_t P=-\sum_{i=1}^{N}\frac{1}{\zeta}\ \ve[\nabla]_i  (\ve[F]_i P)+D\sum_{i=1}^{N}\nabla_i^2 P	\label{FokkerPlanckPositions}
\end{equation}
where $\nabla_i =\partial/\partial \ve[r]_i$ and  $\ve[F]_i$ is the force acting on the $i^{th}$ monomer. This force is related to the monomer positions by  $\ve[F]_i=k(\ve[r]_{i+1}-2\ve[r]_i+\ve[r]_{i-1})$ with the convention that $\ve[r]_0=\ve[r]_1$ and $\ve[r]_{N+1}=\ve[r]_N$. This defines the matrix $M$ such that $\ve[F]_i=-k \sum_{j=1}^{N}M_{ij}\ve[r]_j$. The matrix $M$ is  tridiagonal positive symmetric, its eigenvalues are therefore positive and given by $\lambda_j=2[1-\text{cos}((j-1)\pi/N)]$ with $j \in\{1,...,N\}$. The first eigenvalue is $\lambda_1=0$ and is associated with the motion of the polymer center of mass. The largest relaxation time of the internal conformations of the chain is named the Rouse time $\tau_R=\zeta/(k\lambda_2)=N^2\zeta/(k\pi^2)$.

To cover both cases of intra and intermolecular reactions, we focus on two examples of reactions sketched in Fig. \ref{Fig1sketch}  that are associated to two observables $\ve[R]_{\text{obs}}$. First, the position of the first monomer $\ve[R]_{\text{obs}}=\ve[r]_1$, relevant to intermolecular reactions,  which diffuses at large times with the same  diffusion coefficient $D_{\text{CM}}=D/N$ as the polymer center-of-mass.  Second, the end-to-end vector $\ve[R]_{\text{obs}}=\ve[r]_N-\ve[r]_1$, relevant to intramolecular reactions,  which is not diffusive since  the  mean squared displacement %$\langle\Delta\ve[R]_{\text{obs}}^2\rangle$ 
relaxes in finite time to an equilibrium value. 
To quantify the reaction time for intra and intermolecular reactions, we aim at calculating the mean first passage time \cite{Redner:2001a} for the variable $\ve[R]_{\text{obs}}$ to reach a value $\| {\ve[R]_{\text{obs}}} \|=a$ given an initial probability distribution of the polymer position $P_{\text{ini}}(\vert \ve[r] \rangle)$, where $a$ is the reaction radius (see Fig. \ref{Fig1sketch}). Here $\vert \ve[r] \rangle$ represents the  vector with $N$ components $(\ve[r] _1,...,\ve[r] _N)$ that defines a conformation of the polymer. We will consider the cases where the initial distribution is the stationary distribution $P_{\text{ini}}(\vert \ve[r] \rangle)=P_{\text{stat}}(\vert \ve[r] \rangle)$, or the stationary distribution restricted to values of $\vert \ve[r] \rangle$ such that $\| {\ve[R]_{\text{obs}}} \|=R_{\text{obs}}^0$. In the case of intermolecular reactions, we introduce a large confining volume $V$ in which the reaction takes place (see Fig. \ref{Fig1sketch}b). 
  
While the dynamics of $\ve[R]_{\text{obs}}$ is non Markovian, the evolution of the full polymer conformation $\vert \ve[r] \rangle$ is Markovian and obeys a renewal equation which is the starting point of our analysis.  Let us define $f(\vert \ve[r]\rangle,t)$ the probability density that, starting from the initial distribution, the reactive region %(that is $\| {\ve[R]_{\text{obs}}} \|=a$)%target
 is reached  for the first time at $t$ with a configuration $\vert \ve[r]\rangle$.  
The renewal equation then takes the following form   which is valid for all the configurations $\vert \ve[r]\rangle$ such that $\|\ve[R]_{\text{obs}}\|\le a$:
\begin{equation}
	P(\vert \ve[r]\rangle,t\vert \text{ini},0)=\int_0^{t}dt' \int d\vert \ve[r]'\rangle f(\vert \ve[r]'\rangle,t') P(\vert \ve[r]\rangle,t-t' \vert \ \vert \ve[r]'\rangle,0).\label{renewal}
\end{equation}
Here, $d\vert \ve[r]\rangle=d\ve[r]_1...d\ve[r]_N$,  $P(\vert \ve[r]\rangle,t\vert \text{ini},0)$ is the probability of a configuration $\vert \ve[r]\rangle$ at $t$ in the absence of target when the initial distribution at $t=0$ is $P_{\text{ini}}(\vert \ve[r]\rangle)$, and $P(\vert \ve[r]\rangle,t-t'\vert \ \vert \ve[r]'\rangle,0)$ is the probability of observing the configuration $\vert \ve[r]\rangle$ at $t$ given that the configuration $\vert \ve[r]'\rangle$ was observed at $t=0$. 
We introduce the splitting probability distribution $\pi(\vert \ve[r]\rangle)=\pi(\ve[r]_1,...,\ve[r]_N)$ that represents the probability density of observing a configuration $\vert \ve[r]\rangle$ when the reaction takes place. Taking the Laplace transform of the renewal equation (\ref{renewal}) and developing for small values of the Laplace variable yields an integral equation that links the mean first-passage time $\tau$ and the splitting probability $\pi(\vert \ve[r]\rangle)$:
\begin{align}
	&\tau P_{\text{stat}}(\vert \ve[r]\rangle)=	\int_0^{\infty}dt \left[P(\vert \ve[r]\rangle,t\vert \pi,0)-P(\vert \ve[r]\rangle,t\vert \text{ini},0)\right].	\label{EqIntegraleDim1}
\end{align}
Here, we have introduced $P(\vert \ve[r]\rangle,t\vert \pi,0)$ the probability of a configuration $\vert \ve[r]\rangle$ at $t$ given that the configuration at $t=0$ is taken from the splitting probability $\pi$, which reads:
\begin{equation}
	P(\vert \ve[r]\rangle,t\vert \pi,0)=\int d\vert \ve[r]'\rangle \pi(\vert \ve[r]'\rangle)P(\vert \ve[r]\rangle,t\vert \ \vert \ve[r]'\rangle,0)\label{DefPGivenPi}.
\end{equation}
The equations (\ref{EqIntegraleDim1},\ref{DefPGivenPi}) together with the normalization condition for $\pi(\vert \ve[r]\rangle)$ form an integral equation that completely defines $\pi$ and $\tau$, but which is very difficult to solve in the general case. Let us introduce a final position for the observable $\ve[R]_{\text{obs}}^{\text{f}}$ that is located inside the reactive zone ($\| \ve[R]_{\text{obs}}^{\text{f}}\|\le a$).
Integrating Eq. (\ref{EqIntegraleDim1}) over the $\vert \ve[r]\rangle$ such that $\ve[R]_{\text{obs}}=\ve[R]_{\text{obs}}^{\text{f}}$ gives an exact expression of the mean reaction time  $\tau$ as a function of the splitting probabilities:
\begin{align}	
	\tau P_{\text{stat}}(\ve[R]_{\text{obs}}^{\text{f}})=\int_0^{\infty}dt \Big[P(\ve[R]_{\text{obs}}^{\text{f}},t\vert \pi,0)-P(\ve[R]_{\text{obs}}^{\text{f}},t\vert \text{ini},0)\Big].\label{EstimationMFPT_splitting_dim3}
\end{align}
This exact expression generalizes  the results obtained in Refs. \cite{Condamin2007,Condamin2008,Benichou2010} for Markovian systems. The term $P_{\text{stat}}(\ve[R]_{\text{obs}}^{\text{f}})$ represents the stationary  probability distribution of observing the observable with the value $\ve[R]_{\text{obs}}^{\text{f}}$. In the case of intramolecular reactions, 
this term is sometimes called the \textit{j-factor} \cite{Allemand2006}. In the case of intermolecular reactions, for which  $\ve[R]_{\text{obs}}=\ve[r]_{\text{1}}$, $P_{\text{stat}}(\ve[R]_{\text{obs}}^{\text{f}})=1/V$, where $V$ is the confinement volume, and we will derive below  the large volume  asymptotics of the mean first-passage time. Note that,  by construction, the formula (\ref{EstimationMFPT_splitting_dim3})  provides the same value of $\tau$ for all possible values of $\ve[R]_{\text{obs}}^{\text{f}}$ inside the reactive zone.

The equation (\ref{EstimationMFPT_splitting_dim3}) shows that the calculation of the mean first-passage time requires the determination of the distribution of the polymer configuration at reaction  $\pi$, which is highly non trivial. A local equilibrium assumption, which turns out to give the same results as the Wilemski-Fixman approximation \cite{Pastor1996,WILEMSKI1974a,WILEMSKI1974b},  then consists in approximating the splitting probability by the stationary probability restricted to configurations $\vert \ve[r]\rangle$ such that $\ve[R]_{\text{obs}}$ lies on the surface of the target [$\pi(\vert \ve[r]\rangle)\simeq P_{\text{stat}}(\vert \ve[r]\rangle \vert\ \| \ve[R]_{\text{obs}}\|=a)$]. This approximation is Markovian because it assumes that all the variables relax instantaneously to an equilibrium distribution. Here we go beyond this Markovian assumption and keep track of non Markovian aspects of the problem by calculating $\pi$ in a self-consistent way. As we proceed to show, this distribution $\pi$ of the polymer conformation at the instant of reaction markedly differs from the equilibrium distribution, showing that the non Markovian features of the kinetics cannot be ignored.  For clarity, we present the method in dimension 1, and discuss the results in dimensions 1 and 3. 

In the 1-dimensional case one can take the target size $a=0$ and $x_{\text{obs}}^{\text{f}}=0$ without loss of generality. The key assumption of our approach is that the splitting probability $\pi(\vert x\rangle)=\pi(x_1,...,x_N)$ is a multivariate gaussian distribution that is fully characterized by the averages  (which we call $m^{\pi}_i$) and the covariance matrix (denoted  $\sigma_{ij}^{\pi}$) of the variables $x_i$. The moments of $\pi$ can be calculated by using self-consistent equations, that are derived from (\ref{EqIntegraleDim1}) in the supplementary information (SI):
\begin{align}	
&\int_0^{\infty}dt \left[P(0,t\vert \pi,0)\mu_i^{\pi,0}-P(0,t\vert \text{ini},0)\mu_i^{\text{ini},0}\right]=0
\label{FirstMoment}\\
&\int_0^{\infty}dt \Big[P(0,t\vert \pi,0)\left(\gamma_{ij}^{\pi,*}+\mu_i^{\pi,0}\mu_j^{\pi,0}-\sigma_{ij}^{\text{stat},*}\right)
-P(0,t\vert \text{ini},0)\left(\gamma_{ij}^{\text{ini},*}+\mu_i^{\text{ini},0}\mu_j^{\text{ini},0}-\sigma_{ij}^{\text{stat},*}\right)\Big]=0
	\label{2ndMoment}
\end{align}
Here, $\mu_i^{\pi,0}$ and  $\gamma_{ij}^{\pi,*}$ are the moments of the distribution $P(\vert x\rangle,t\vert x_{\text{obs}}=0,t;\pi,0)$ (the distribution of the configurations $\vert x\rangle$ at $t$  such that $x_{\text{obs}}=0$ and given that the initial configuration is taken from the splitting probability). These moments $\mu_i^{\pi,0}$ and  $\gamma_{ij}^{\pi,*}$ can be related to the initial moments $m_i^{\pi}$ and  $\sigma_{ij}^{\pi}$ by  projection and propagation formulas that are given in  SI. In the same way, $\mu_i^{\text{ini},0}$ and $\gamma_{ij}^{\text{ini},*}$ are the moments of the distribution $P(\vert x\rangle,t\vert x_{\text{obs}}=0,t;\text{ini},0)$, whereas $\sigma_{ij}^{\text{stat},*}$ is the covariance matrix of the distribution $P_{\text{stat}}(\vert x\rangle\vert x_{\text{obs}}=0)$. Last, the propagator $P(0,t\vert \pi,0)$ can be written  as a function of $m_i^{\pi}$ and  $\sigma_{ij}^{\pi}$ (see SI). Equations (\ref{FirstMoment},\ref{2ndMoment}) provide a closed system of explicit equations that fully determines  the unknown moments $m_i^{\pi}$ and  $\sigma_{ij}^{\pi}$, and therefore the distribution $\pi(\vert x\rangle)$. These equations can be extended to the case of a 3-dimensional space (see SI). Together with Eq. (\ref{EstimationMFPT_splitting_dim3}), Eqs.(\ref{FirstMoment},\ref{2ndMoment}) finally enable the determination of the mean reaction time and constitute one of the main results of our work. 

\vspace{1cm}
{\bf Discussion}

\begin{figure}[h!] 
\includegraphics[width=8cm,clip]{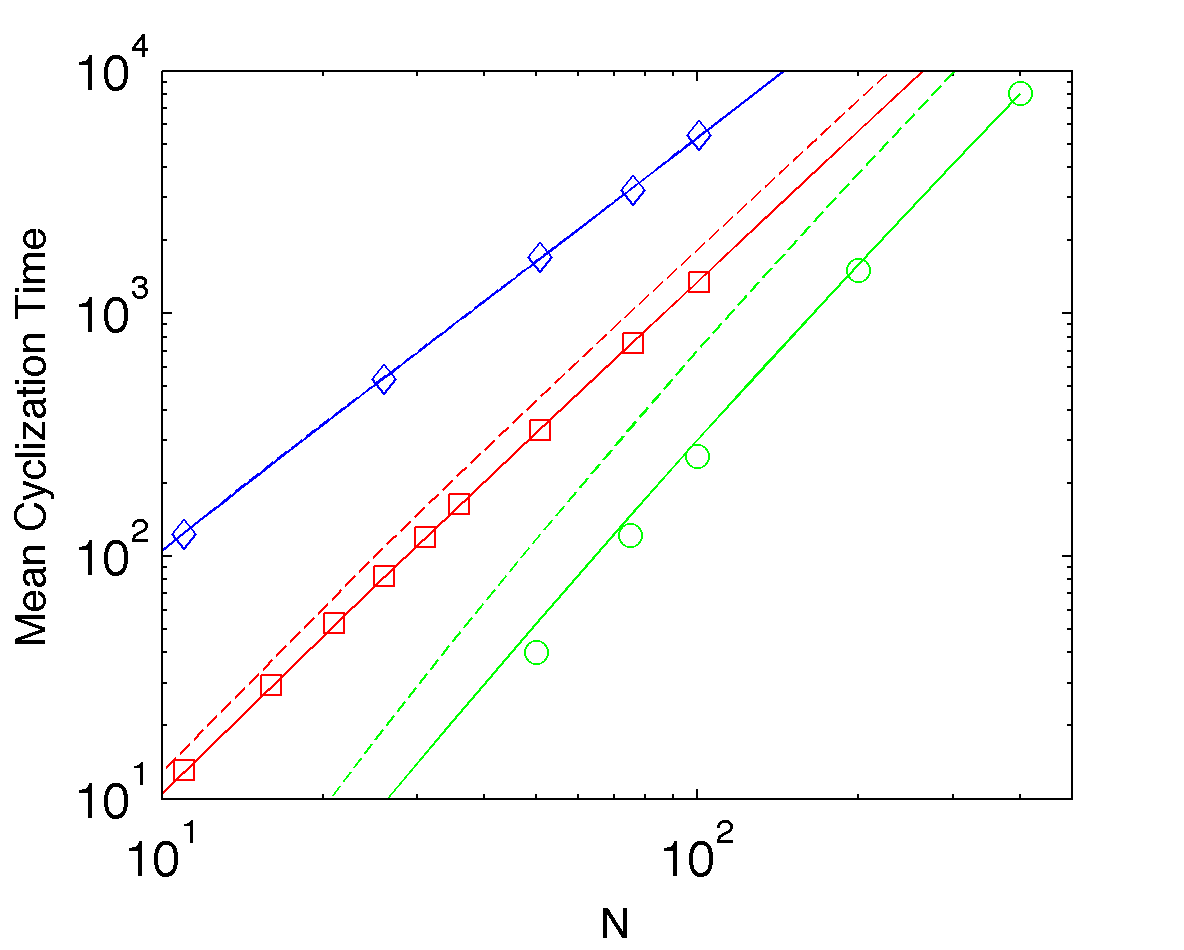}   
 \caption{{\bf Theory and simulations of the cyclization reaction}. Comparison between the markovian theory (dashed lines), the non-markovian theory (continuous lines) and simulation results (symbols). Different colors correspond to different sizes of reactive region (blue: $a/l_0=\sqrt{3}/10$, red: $a/l_0=\sqrt{3}$, green: $a/l_0=5\sqrt{3}$). The squares and diamonds symbols correspond to the simulations data of Ref. \cite{Pastor1996}, whereas the circles correspond to our Brownian dynamics simulations. The non-markovian results are calculated by approximating the covariance matrix of the splitting probability by its stationary value. The units of length and time are $l_0$ and $\zeta/k$.}
\label{FigCyclization}
\end{figure}

\begin{figure}[h!]
 \includegraphics[width=12cm,clip]{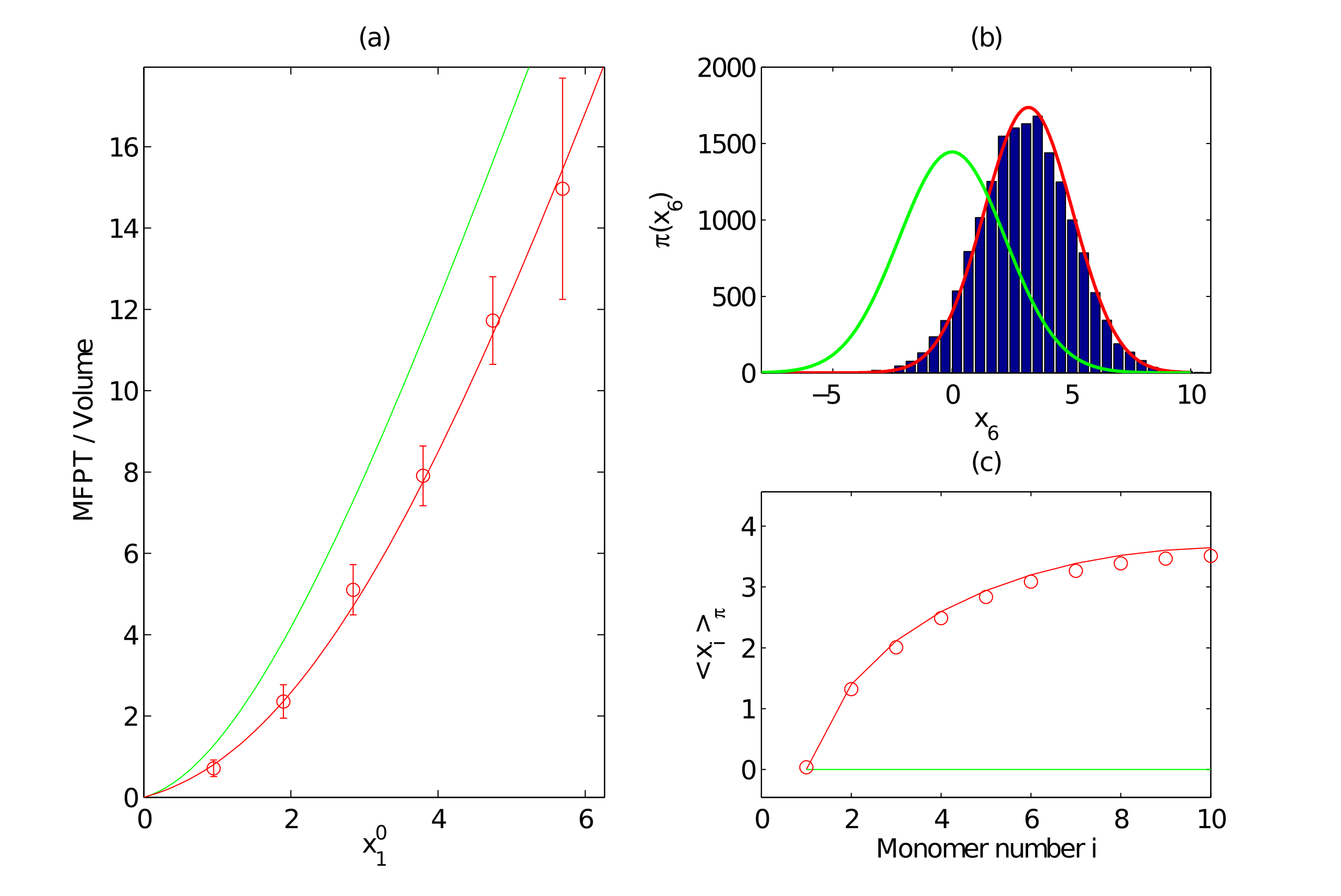}   
 \caption{{\bf Theory and simulations of a diffusive variable in one dimension: reaction of the first monomer of a Rouse chain with $N=10$ monomers with a target located at $x=0$ in a confining volume $V$.} (a) Mean first passage time (rescaled by the volume) as a function of the initial distance $x_1^0$. Red symbols: results of numerical simulations for a volume $V=316l_0$, error-bars are $95\%$ confidence intervals. Lines: theoretical estimations of $\tau/V$ (green: markovian approximation ; red: non markovian theory). (b) Marginal splitting probability histogram for the $6^{\text{th}}$ monomer position ($\pi(x_6)$) when the starting position is $x_1^0=5.7 l_0$. Lines: theoretical distribution predicted by the non Markovian theory (red) and the Markovian approximation (green). (c) Average position of the monomers at the reaction for $x_1^0=5.7 l_0$ (simulation and theory, with the same colors and symbols as in (a)). The units of length and time are $l_0$ and $\zeta/k$.}
\label{temps1d}
\end{figure}

\begin{figure}[h!]
 \includegraphics[width=12cm,clip]{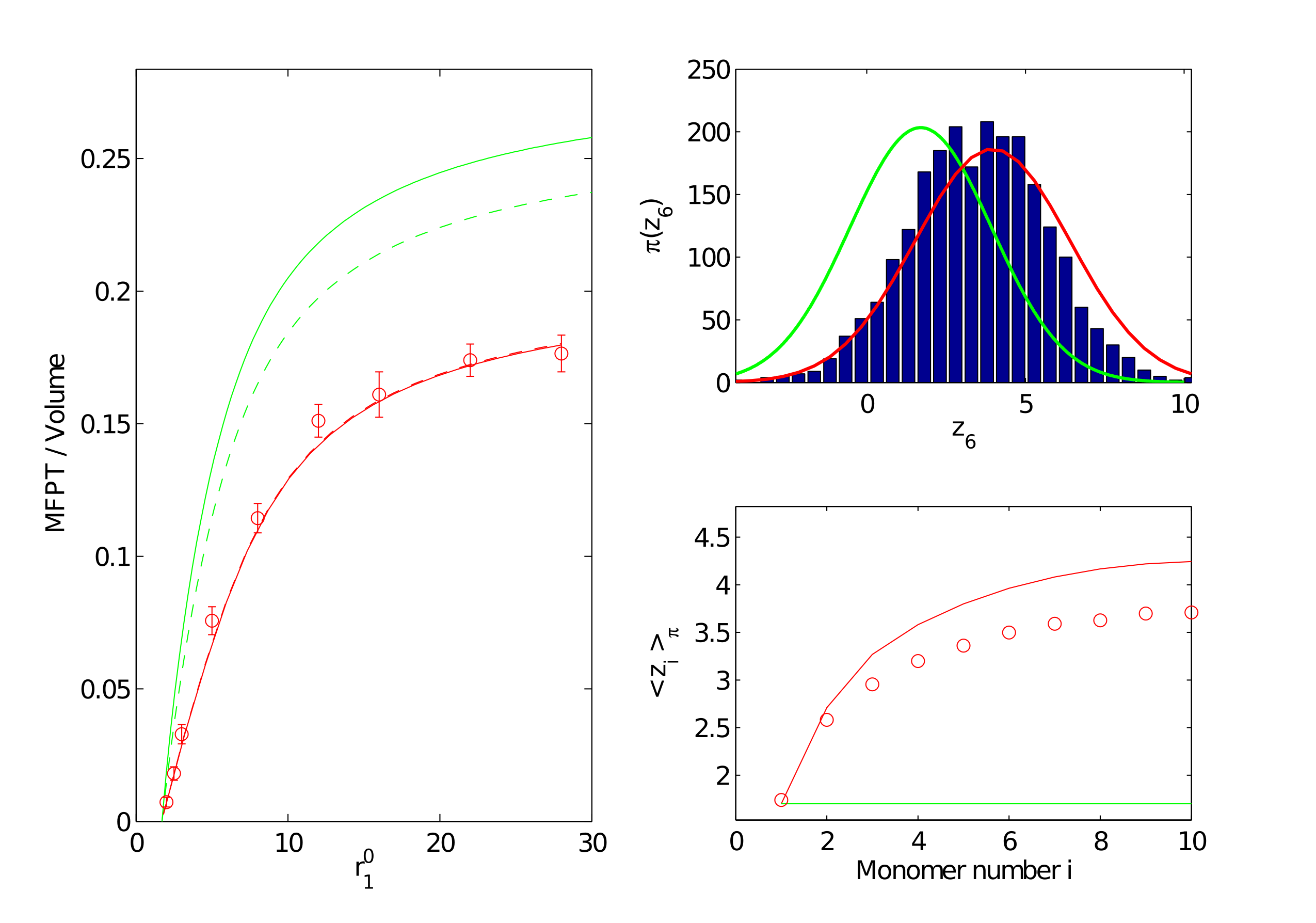}   
 \caption{{\bf Theory and simulations of a diffusive variable in 3 dimension: reaction of the first monomer of a Rouse chain with $N=10$ monomers with a target of size $a=1.7l_0$  centered around the position $\ve[r]=0$ in a confining spherical volume $V$ of radius $28.5 l_0$}. The color code is the same as in Fig \ref{temps1d}. (a) Mean first reaction time $\tau(r_1^0)$ (rescaled by the confining volume) as a function of the initial distance between the reactants. Symbols: simulations ; Lines: theoretical estimations (green: markovian approximation ; red: non-markovian theory).
Dashed lines give estimations of $\tau$ by using $\|\ve[R]_{\text{obs}}^{\text{f}}\|=a$, whereas the continuous lines are calculated with $\ve[R]_{\text{obs}}^{\text{f}}=0$. All calculations assume the spatial isotropy of the covariance matrix. (b) Bars: histogram of the radial component of the position of the $6^{\text{th}}$ monomer at the reaction when $r_1^{0}=28l_0$. Lines: theoretical distribution predicted by the non Markovian theory (red) and the Markovian approximation (green). (c) Radial average positions of the monomers at the reaction when $r_1^{0}=28l_0$ (red symbols), compared with the theoretical predictions of the markovian theory (green line) and of the non-markovian theory (red line). The units of length and time are $l_0$ and $\zeta/k$.}
\label{temps3d}
\end{figure}

In order to test the validity of our non Markovian approach, we compared its predictions with the results of several  numerical  simulations in  1 and 3 dimensions for both cases of inter and intramolecular reactions (see Figs \ref{FigCyclization},\ref{temps1d},\ref{temps3d}). The Brownian dynamics simulations were carried out by using the algorithms presented in Refs. \cite{Pastor1996,Peters2002} (see SI for details).
 Since the Wilemski-Fixman approach is the Markovian theory  that matches best  simulation results until now \cite{Pastor1996}, we also compare our predictions to this theory, which will be referred to below as    the Markovian theory.

In the case of intramolecular reactions exemplified by the cyclization reaction (see Fig. \ref{Fig1sketch}a), our non Markovian theory is in excellent agreement with the simulations for all sizes $a$ of the reactive region and for all polymer lengths, and significantly improves the results of the Markovian theory (Fig \ref{FigCyclization}). To our knowledge, our non Markovian approach is the first theory that provides an accurate description of the cyclization time for all ranges of parameters.
 In the case of intermolecular reactions where for example the reactive site is the  first monomer of a chain (see Fig. \ref{Fig1sketch}b), our theory yields an excellent quantitative determination of the reaction time and significantly improves the results of existing Markovian theories (see Figs. \ref{temps1d}a,\ref{temps3d}a). As stated above, the key element of our analysis is the determination of the distribution of the polymer conformation at the very instant of reaction. 
The very precise  determination of the mean first passage time is a consequence of the fact that the Gaussian approximation accurately describes the distribution $\pi$ (see Figs \ref{temps1d}b,c and Figs \ref{temps3d}b,c). We stress that the Gaussian prediction of our non Markovian approach markedly differs from the equilibrium distribution implicitly assumed in Markovian theories. Remarkably,  the reactive conformation of the polymer for both inter and intramolecular reactions is actually significantly more extended than the equilibrium conformation, which yields reaction times notably shorter than predicted by Markovian theories.

We now derive  analytical formulas for  the mean reaction time from Eqs. (\ref{EstimationMFPT_splitting_dim3},\ref{FirstMoment},\ref{2ndMoment}) in different limiting regimes in the most relevant case of dimension 3. These  simple explicit formulas  enable the definition of the different regimes of polymer reaction kinetics, and establish clearly the validity  domains of the Markovian theories. 
Scaling relations rely in part on the behavior of the  mean squared displacement of a monomer with time:
\begin{align}
\langle\Delta\ve[r]^2\rangle=
\begin{cases}
	6 D t & \text{if } t\ll \zeta/k \\
	\alpha t^{1/2} & \text{if }\zeta/k\ll t\ll \tau_R\sim N^2\zeta/k\\
	6 D_{\rm CM} t & \text{if } t\gg  \tau_R
\end{cases}\label{ScalingBehavior}
\end{align}
where $\alpha$ is a numerical coefficient. The anomalous diffusion at intermediate time scales involves all the time scales of the Rouse chain and  becomes important  in the  limit of long chains  $N\gg 1$.  At longer time scales, the behavior of the monomer is diffusive with diffusion coefficient $D_{\rm CM}$. 

\textit{Small target regime $a\ll l_0/\sqrt{N}$}. 
In this limit of a  target much smaller than the bond length, the reaction kinetics depends essentially on the short time properties of the search, and the mean reaction time  can be shown from Eqs. (\ref{EstimationMFPT_splitting_dim3},\ref{FirstMoment},\ref{2ndMoment}) to asymptotically follow:
\begin{align}
	\tau 
&=\begin{cases}
V/(4\pi D a ) & \text{(intermolecular reaction)}\\
      \sqrt{\pi/8}(l_0 N)^{3/2}/(Da)  & \text{(intramolecular reaction)}.
\end{cases}		 \label{tauEquivalentSSS}
\end{align}
This shows in particular that for large $N$ the mean cyclization time scales as $\tau\sim N^{3/2}/a$ for small targets. The strong dependance of the mean reaction time with the target size in this limit is the signature of a non-compact exploration \cite{DEGENNES1982}. 
It is notable that in this non-compact limiting case, the non Markovian theory predicts the same result as the two classical Markovian theories (the Wilemski-Fixman theory \cite{WILEMSKI1974a,WILEMSKI1974b} and the harmonic spring model \cite{Szabo1980,Pastor1996,SUNAGAWA1975}) in the limit of small reaction radius, which validates Markovian approaches in this regime.

\begin{figure}[h!]
 \includegraphics[width=14cm,clip]{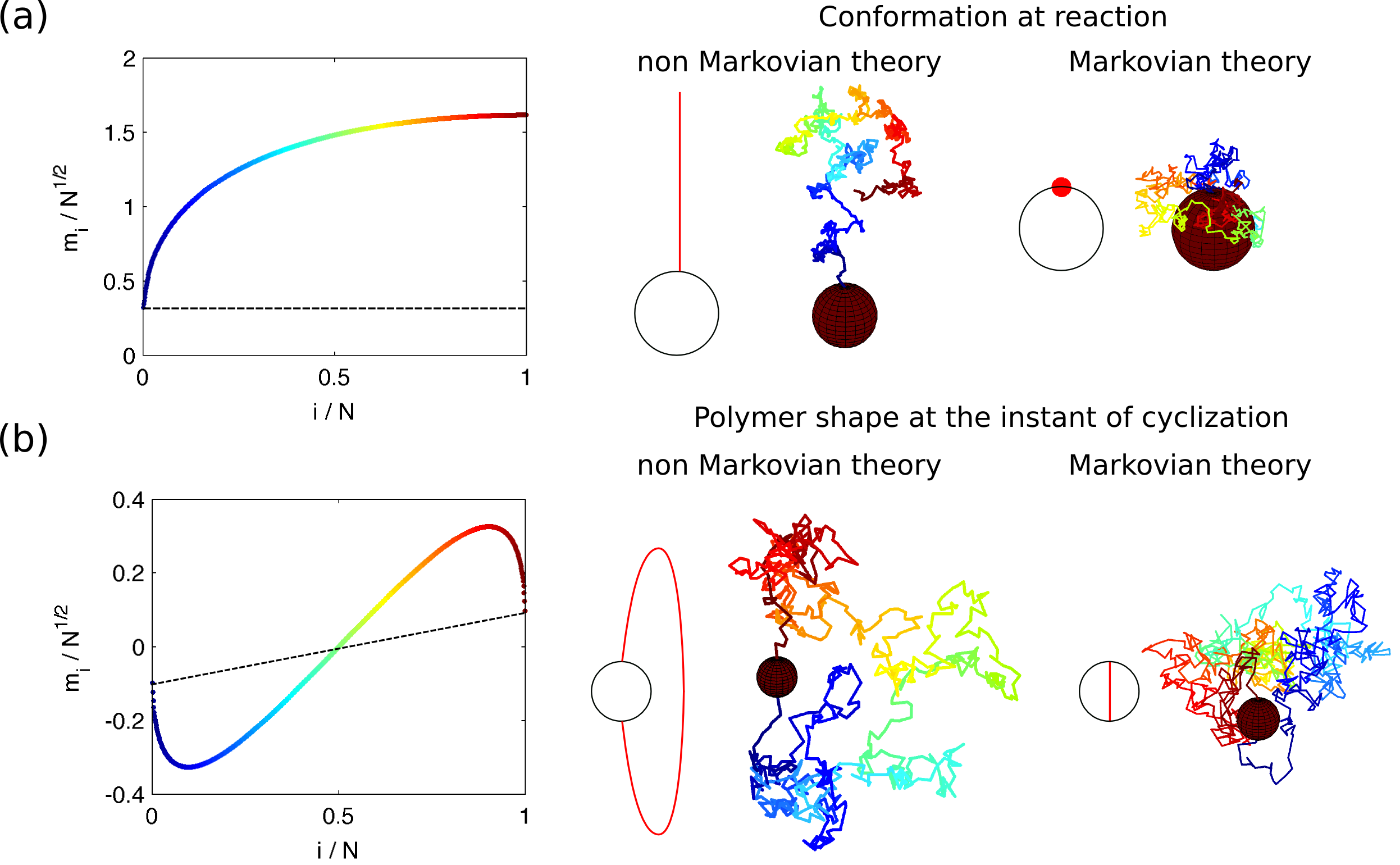}   
 \caption{{\bf Predicted polymer conformations at reaction.} Left: Average radial position of the monomers when the reaction takes place for: (a) the reaction between the first monomer and a target  and: (b) the cyclization reaction  (Continuous line: prediction of non Markovian theory, dashed line: Markovian approximation). For both reactions, we also plot  the sketch of the polymer shape when the reaction takes place, and an example of conformation drawn from the splitting probability distribution (left) which is in marked contrast with  the stationary distribution (right). The reaction is assumed to take place along the vertical axis. The position of a monomer in the chain is represented by a color code. The sketch of the polymer shape for the cyclization reaction  is artificially extended in the horizontal direction for clarity. Parameter values: (a): $N=800$ and $a=0.32l_0\sqrt{N}$; (b): $N=800$ and $a=0.094l_0\sqrt{N}$. The unit of length is $l_0$.}
\label{FormeConfiguration}
\end{figure}

%Thermodynamic limit
\textit{Intermediate target regime $l_0/\sqrt{N}\ll a< l_0\sqrt{N}$}.  In the regime of long polymer chains, a single monomer displays at intermediate time scales a subdiffusive behavior $\langle \Delta \ve[r]^2\rangle\sim t^{2/d_w}$ [Eq. (\ref{ScalingBehavior})], thereby defining a walk dimension  $d_w=4$ that is larger than the spatial dimension $d=3$. As a consequence, a monomer is able to densely explore the space, and the time to reach a target much smaller than  the polymer size $l_0\sqrt{N}$ is asymptotically independent of the target size $a$. In the case of intermolecular reactions, we show from Eqs. (\ref{EstimationMFPT_splitting_dim3},\ref{FirstMoment},\ref{2ndMoment}) that the reaction time averaged over all initial conditions reads:
\begin{align}
	\tau =\frac{V}{4\pi D_{\text{CM}}a_{\text{eff}}(a,N)} \label{GlobalMFPTDIffusif}
\end{align}
where $a_{\text{eff}}(a,N)$ is an effective target size, which is of the order of the polymer size ($a_{\text{eff}}\sim l_0 \sqrt{N}$) and  does not depend on the real size of the target $a$ for  $a\rightarrow0$. This equation has a clear interpretation : the polymer only needs to approach the target at a distance comparable to $l_0\sqrt{N}$, and then the reaction takes place instantaneously due to the compact search at small length scales. Importantly, if both the early analysis of De Gennes \cite{DEGENNES1982} and the Markovian theory predict the same functional form (\ref{GlobalMFPTDIffusif}), they fail  to predict a correct estimate of the effective target size $a_{\text{eff}}$, which is found in realistic regimes to be underestimated by a factor of two as compared to the non Markovian result (see SI).
This difference can be understood from the fact that the polymer is much more extended at the  instant of  reaction than in its equilibrium conformation (Fig. \ref{FormeConfiguration},a). %It is remarkable that this extension is in the direction defined by the entrance point in the reactive region, even in the limit of small target.
 In the non Markovian description, the polymer center-of-mass therefore needs to approach  the target less closely than in the Markovian theory, leading to a faster reaction kinetics. 
Similarly, we find in the case of intramolecular reactions that the mean cyclization time (averaged over  stationary initial configurations) is given by
\begin{align}
	\tau =c\left(\frac{a}{l_0\sqrt{N}}\right)\frac{ N^2 \zeta}{ \pi^2k},
	\label{GlobalCyclizationTime}  
\end{align}
where again the numerical function $c(a/l_0\sqrt{N})$ is typically underestimated by a factor of two by the Markovian theory (see SI).

The faster kinetics predicted by the non Markovian theory for both intra and intermolecular reactions is a direct consequence of  the non trivial out of equilibrium distribution of the polymer conformation at the instant of reaction (see Fig. \ref{FormeConfiguration}). While the Markovian theory implicitly assumes an equilibrium conformation, the non Markovian theory shows that the reaction takes place when the polymer is in fact much more extended than in its  equilibrium conformation, thus increasing the effective reaction radius, as seen in Fig. \ref{FormeConfiguration},b. This contribution of non Markovian effects turns out to be quantitatively important, since in this regime of intermediate targets the Markovian approximation leads to an error in the estimate of the reaction time of roughly $100\%$.

While we have here focused on the Rouse model,  we stress that the fact that such non Markovian effects are characterized by non equilibrium  polymer conformations at the instant of the reaction holds true for more general models of polymer dynamics. Our method can  in fact be extended   to  general gaussian models, which play a key role in polymer dynamics and enable the modeling of various physico-chemical conditions. Examples of such gaussian theories include the ``pre-averaging'' approach of hydrodynamic interactions \cite{DoiEdwardsBook}, 
the approximate Rouse modes in the case of self-avoiding polymers \cite{Panja2009}, and the description of semi-flexible chains and branched polymers with a gaussian theory \cite{Dua2002,Dolgushev2011}.  
 Qualitatively, we expect the non Markovian effects to be significant when the search at small time scales is compact, so that the transport step plays a crucial role in the kinetics. This includes the case of self-avoiding chains as well as chains with hydrodynamic interactions in theta solvent.
Last, at the experimental level, the formation of hairpins in nucleic acids \cite{Bonnet1998,Wallace2001} or the folding of polypeptide chains \cite{Lapidus2000} constitute important examples of cyclization reactions. So far, only Markovian theories of such reactions have been used to interpret observations \cite{Wallace2001,Lapidus2000} and we anticipate that  taking into account non Markovian effects as quantified by our approach could improve the quantitative analysis of experimental data.

To conclude, we proposed a new theory of polymer reaction kinetics that takes into account the non Markovian effects that control the dynamics of polymers. This non Markovian theory gives results that are in quantitative agreement with numerical simulations for all  ranges of parameters, and therefore significantly outperforms existing Markovian approaches.  
%Our theory provides a physical interpretation that leads to the difference observed between Markovian and non Markovian approaches. 
Our analysis reveals that the non equilibrated conformation of the polymer at the instant of the reaction has an important impact on the reaction kinetics. We show quantitatively that the typical reactive conformation of the polymer is more extended than the equilibrium conformation, leading to reaction times that are significantly shorter than predicted by existing Markovian theories. Together, our results provide a better understanding of the complex kinetics of polymer reactions  involved for example in the formation of loops of RNA or polypeptides chains.

\vspace{1cm}
{\bf Supplementary information}
\appendix
\section{Derivation of the self-consistent equations (5,6,7) of the main text.}
\subsection{Definition of the Rouse modes and choice of units}
Before describing how to obtain the self-consistent equations (5,6,7) of the main text, we introduce the notion of the Rouse modes, that considerably simplifies the Fokker-Planck equation (1). We remind that the matrix $M_{ij}$ that links the forces on the monomers $\ve[F]_i$ to the positions $\ve[r]_j$ is defined by the relation: $\zeta \ve[F]_i=- k \sum_{j=1}^{N} M_{ij}\ve[r]_j$. $M$ is therefore the following $N\times N$ tridiagonal matrix:
\begin{equation}
M=\begin{pmatrix}
	 1 & -1 & 0 & .. & .. &..&..\\
	-1 & 2 & -1 & 0 & .. & ..&.. \\
	0 & -1 & 2 & -1 & 0 & ..&.. \\
		.. & .. &.. & .. & .. & ..&.. \\
	 ..&.. & 0& -1 & 2 & -1&0\\
			..& .. & .. & 0& -1 & 2 & -1 \\
..& ..&..&..&0& -1 & 1\\	
	\end{pmatrix}
\end{equation}
Because $M$ is symmetric positive, it can be diagonalized. The eigenvalues of $M$ are:
\begin{align}
	\lambda_j=2\{1-\text{cos}[(j-1)\pi/N]\}\label{ValeurDesValeursPropres}
\end{align}
We can write $M=QDQ^{-1}$, with $D$ the diagonal matrix with $(\lambda_1,...,\lambda_N)$ on the diagonal and the passage matrix $Q$ is:
\begin{align}
	Q_{i1}=1/\sqrt{N} \ ;Ê\	Q_{ij}=\sqrt{\frac{2}{N}}\text{cos}\left[(i-1/2)(j-1)\pi/N\right] \ \ \text{if} \ \  j\ge2 \label{DefinitionMatrixQ}
\end{align}
Note that the inverse of $Q$ is its transpose matrix. We define the Rouse modes $(\ve[a]_1,...,\ve[a]_N)$ by the relations:
\begin{equation}
	\ve[r]_i=\sum_{j=1}^{N} Q_{ij}\ve[a]_j  \ \ ; \ \ \ve[a]_i=\sum_{j=1}^N Q_{ji}\ve[r]_j		\label{DefinitionModes}
\end{equation}
Note that the transformation (\ref{DefinitionModes}) is equivalent to the Fourier transform when $N$ is large. We study the observables $\ve[R]_{\text{obs}}$ that can be expressed as a linear combination of the positions $\ve[r]_i$, or equivalently of the Rouse modes $\ve[a]_j$. An observable $\ve[R]_{\text{obs}}$ is therefore associated with a set of $N$ coefficients $(b_1,...,b_N)$:
\begin{align}
	\ve[R]_{\text{obs}}=\ve[R]=\sum_{i=1}^N b_i \ve[a]_i = \langle b \vert \ve[a]\rangle \label{DefinitionObservable}
\end{align}
In order to have more contracted notations, we omit the subscript of $\ve[R]_{\text{obs}}$ and simply call the observable $\ve[R]$ (we reserve capital letter for the observable).
In equation (\ref{DefinitionObservable}), we have introduced the notation that $\vert u\rangle$ is a column vector with $N$ components $(u_1,...,u_N)$, $\langle u\vert$ is its transpose, and $\langle u\vert v\rangle=\sum_{i=1}^N u_i v_i$ is the scalar product between the vectors $\vert u\rangle$ and $\vert v\rangle$. Note that quantities in bold represent vector in the physical 3-dimensional space, to be distinguished from the $N$ components vectors noted $\vert u\rangle$. If the observable is the end-to-end vector ($\ve[R]=\ve[r]_N-\ve[r]_1$), then the coefficients are $b_i=Q_{Ni}-Q_{1i}$. In the case that the observable is the position of the first monomer ($\ve[R]=\ve[r]_1$), the coefficients are $b_i=Q_{1i}$. We distinguish the \textit{diffusive observables} (for which $b_1\ne0$, these observables diffuse as the polymer center--of--mass at large times) from the \textit{non-diffusive observables} (for which $b_1=0$). Finally, in all the supplementary information, we choose the units such that $D=1$, $\zeta=1$ and $k=1$. The unit of energy is $k_BT=1$, the unit of time is $\zeta/k$, and the unit of length is the effective bond length $l_0=\sqrt{k_BT/k}=1$. In these units, the effective Kuhn length is $l_{\text{Kuhn}}=\sqrt{3}$. 

\subsection{Non Markovian theory in dimension $d=1$}
\subsubsection{Self-consistent equations for the non Markovian theory in dimension $d=1$}
We now describe how to derive the equations (5,6,7) of the main text that define the non Markovian theory in dimension $d=1$. The observable is then noted $X$, and we calculate the mean first passage time for $X$ to reach the value $X=0$.
The starting point is Eq. (3) (in the main text), which can be written in terms of modes:
 \begin{align}
	&\tau P_{\text{stat}}(\vert a\rangle)=	\int_0^{\infty}dt \left[P(\vert a\rangle,t\vert \pi,0)-P(\vert a\rangle,t\vert \text{ini},0)\right]	
\end{align}
This equation is valid only for the modes $\vert a\rangle$ such that $\langle b\vert a\rangle=0$. We note the mathematical trick: $P_{\text{stat}}(\vert a\rangle)\delta(\langle b\vert a\rangle-X)=P_{\text{stat}}(X)P_{\text{stat}}(\vert a\rangle \vert \langle b \vert a\rangle =X)$. Therefore, multiplying Eq. (\ref{EqIntegraleDim1}) by $\delta(\langle b\vert a\rangle)$ leads to a reinterpretation:
 \begin{align}
	\tau P_{\text{stat}}(0) P_{\text{stat}}(\vert a\rangle \vert 0)=	
	\int_0^{\infty}dt [P(0,t \vert \pi,0)P(\vert a\rangle,t\vert 0,t ; \pi,0)
-P(0,t\vert \text{ini},0)P(\vert a\rangle,t\vert 0,t ; \text{ini},0)]	\label{EqIntegraleDim1Reinterpreted}
\end{align}
where  $P(\vert a\rangle,t\vert 0,t ; \pi,0)$ is the probability of observing the configuration $\vert a\rangle$ at $t$ given that the value of the observable $X$ is $X=0$ at $t$ and that the distribution of modes at $t=0$ was $\pi$. Similarly, $P_{\text{stat}}(\vert a\rangle \vert 0)$ is the stationary probability to observe a configuration given that the value of the observable is $X=0$. Now, the equation (\ref{EqIntegraleDim1Reinterpreted}) is valid for any value of $\vert a\rangle$ (not only for those that satisfy $\langle b\vert a\rangle=0$). Noting that the distribution $P(\vert a\rangle,t\vert X,t ; \pi,0)$ is normalized to $1$, the integration of Eq. (\ref{EqIntegraleDim1Reinterpreted}) over all the modes leads to:
 \begin{align}
	\tau P_{\text{stat}}&(0) =	\int_0^{\infty}dt [P(0,t \vert \pi,0)-P(0,t\vert \text{ini},0)]\label{EstimationTauDim1}
\end{align}
Then, multiplying Eq. (\ref{EqIntegraleDim1Reinterpreted}) by $a_i$ and integrating over all the modes leads to:
 \begin{align}
	\int_0^{\infty}dt \left[P(0,t \vert \pi,0)\mu_i^{\pi,0}-P(0,t\vert \text{ini},0)\mu_i^{\text{ini},0}\right]=0 \label{FirstMomentAppendice}
\end{align}
where $\mu_i^{\pi,0}$ is the mean value of $a_i$ at $t$ given that $X=0$ at $t$ and that the initial distribution at $t=0$ is the splitting distribution $\pi$. Similarly, multiplying Eq. (\ref{EqIntegraleDim1Reinterpreted}) by $a_i a_j$, integrating it over all the modes and using Eq. (\ref{EstimationTauDim1}) leads to a second set of self consistent equations:
\begin{align}
\int_0^{\infty}dt \Big[&P(0,t\vert \pi,0)\left(\gamma_{ij}^{\pi,*}+\mu_i^{\pi,0}\mu_j^{\pi,0}-\sigma_{ij}^{\text{stat},*}\right)-P(0,t\vert \text{ini},0)\left(\gamma_{ij}^{\text{ini},*}+\mu_i^{\text{ini},0}\mu_j^{\text{ini},0}-\sigma_{ij}^{\text{stat},*}\right)\Big]=0	\label{2ndMomentAppendice}
\end{align}
where $\gamma_{ij}^{\pi,*}$ is the covariance between $a_i$ and $a_j$ at $t$ given that $X=0$ at $t$ and that the initial distribution at $t=0$ is the splitting distribution $\pi$.
We now derive ``propagation'' and ``projection'' formulas that will be useful to explicitly write all the terms appearing in Eqs. (\ref{EstimationTauDim1},\ref{FirstMomentAppendice},\ref{2ndMomentAppendice}).

\subsubsection{Propagation and projection formulas}
The Fokker-Planck equation that governs the evolution of the Rouse modes (in one dimension) is:
\begin{equation}
	\frac{\partial P(\vert a\rangle,t)}{\partial t}=\sum_{i=1}^{N} \frac{\partial}{\partial a_i}\left(\lambda_i a_iP+\frac{\partial}{\partial a_i}P\right)\label{FKPRouseModes}
\end{equation}
It is well known that Eq. (\ref{FKPRouseModes}) admits Gaussian solutions \cite{VanKampen1992} which are characterized by the average $\mu_i$ of each mode $a_i$ and the covariance matrix $\gamma_{ij}$  that describes the correlations between the modes $a_i$ and $a_j$. The evolution of $\mu_i$ and $\gamma_{ij}$ satisfies the following equations \cite{VanKampen1992} that are sometimes called  generalized fluctuation-dissipation relations (see \cite{FOX1978}):
\begin{align}
	\dot{\mu}_i=-\lambda_i \mu _i Ê\ ; \  \dot{\gamma}_{ij}=-(\lambda_i+\lambda_j)\gamma_{ij}+2\delta_{ij} \label{EqGammaIJ}
\end{align}
If the initial condition is a gaussian distribution with moments $m_i$ and $\sigma_{ij}$, then the value of $\mu_i$ and $\gamma_{ij}$ is the solution of Eq. (\ref{EqGammaIJ}) with initial conditions $\mu_i(0)=m_i$ and $\gamma_{ij}(0)=\sigma_{ij}$. We find:
\begin{align}
	\mu_i& =m_i \ e^{-\lambda_i t} \label{PropagationMean}\\
	\gamma_{ij} &= \delta_{ij}\left(1-e^{-2\lambda_i t}\right)/\lambda_i+e^{-\lambda_i t}e^{-\lambda_j t}\sigma_{ij}\label{PropagationCovariance}
\end{align}
These formulas describe how the mean vector and the covariance matrix are modified with time, and we call them ``propagation formulas''. Note that Eq. (\ref{PropagationCovariance}) is written with the convention that $(1-e^{-2\lambda_1t})/\lambda_1=2t$ (with $\lambda_1=0$).

Let us now assume that $P(\vert a\rangle)$ is a gaussian distribution, with means $\mu_i$ and covariance matrix $\gamma_{ij}$. We describe how to obtain the distribution $P(\vert a\rangle \vert X)$, which represents the distribution of modes given that the value of the observable is $X$. Noting that $P(X\vert\  \vert a\rangle )=\delta(\langle b\vert a\rangle-X)$, and using the Bayes formula, we obtain:
\begin{equation}
	P(\vert a\rangle \vert X)=\delta\left(\langle b\vert a\rangle -X\right)P(\vert a\rangle)/P(X)\label{DefMarginalRestricted}
\end{equation}
We note that the distribution $P(X)$ is gaussian, with means $\langle b\vert a\rangle$ and variance $\langle b\vert \gamma\vert a\rangle$. The distribution $P(\vert a\rangle \vert X)$ is also a gaussian distribution, and we call $\mu_i^{X}$ and $\gamma_{ij}^X$ its vector and covariance matrix. By definition,  we can identify the value of  $\mu_i^{X}$ by writing:
\begin{align}
	\mu_i^{X}&=\int d\vert a\rangle \  a_i \ P(\vert a\rangle \vert X)\label{DefinitionMuIStar}\\
	&=\mu_i-\frac{\langle  e_i\vert\gamma\vert b \rangle }{\langle  b\vert\gamma\vert b \rangle }(\langle b\vert \mu \rangle -X)\label{ProjectionMean}
\end{align}
The passage from Eq. (\ref{DefinitionMuIStar}) to (\ref{ProjectionMean}) results from the explicit calculation of the integral (\ref{DefinitionMuIStar}) with the use of Eq. (\ref{DefMarginalRestricted}), and $\vert e_i\rangle$ represents the $i^{\text{th}}$ basis vector (all its elements are 0 except for the $i^{\text{th}}$ which takes the value $1$). The elements of the covariance matrix can  be identified with the same method:
\begin{align} 
	\gamma_{ij}^{X}&=\int d\vert a\rangle \  (a_i-\mu_i^{X})(a_j-\mu_j^{X}) \ P(\vert a\rangle \vert X)\label{DefinitionGammaIJStar}\\	
	&=\gamma_{ij}-\frac{\langle  e_i\vert\gamma\vert b \rangle \langle  e_j\vert\gamma\vert b \rangle }{\langle  b\vert\gamma\vert b \rangle } =	\gamma_{ij}^{*}\label{ProjectionCovariance}
\end{align}
Note that $\gamma^X$ does not depend on the value of $X$, which is why we just note it $\gamma^*$ and not $\gamma^{X}$, to the difference of $\mu^{X}$ which depends linearly on the value of $X$.
We call the equations (\ref{ProjectionMean}) and (\ref{ProjectionCovariance}) ``projection formulas'': they describe how the mean and the covariance of the $a_i$ are modified when one restricts the modes to be on the hyperplane of equation $\langle b\vert a\rangle =X$. We can easily see that the average of $X$ over the distribution $P(\vert a\rangle \vert X)$ is $\langle b\vert \mu^X\rangle=X$, and that the covariance of $X$ vanishes (because $\gamma^*\vert b\rangle=\vert 0\rangle$), which means that $X$ is known with certainty to be $X$ on this distribution, as expected from the definition (\ref{DefMarginalRestricted}). 

\subsubsection{Explicit expressions of all the terms appearing in the self-consistent equations}
We remind that $m_i^{\pi}$ and $\sigma_{ij}^{\pi}$ are the mean vector and covariance matrix of the splitting distribution $\pi(a_1,...,a_N)$. The average $\mu_i^{\pi}$ and covariance $\gamma_{ij}^{\pi}$ of the modes $a_i$ at $t$ starting from $\pi$ at $t=0$ are deduced from the ``propagation formulas'' (\ref{PropagationMean},\ref{PropagationCovariance}):
\begin{align}
	\mu_i^{\pi} =m_i^{\pi} \ e^{-\lambda_i t} \ ; \ 
	\gamma_{ij}^{\pi} = \delta_{ij}(1-e^{-2\lambda_i t})/\lambda_i+e^{-\lambda_i t}e^{-\lambda_j t}\sigma_{ij}^{\pi} \label{PropagationPi}
\end{align}
Noting that $\langle b\vert \mu^{\pi} \rangle$ and $\langle b\vert \gamma^{\pi} \vert b\rangle$ are the average value of $X$ and the covariance of $X$ at $t$ starting from the initial distribution $\pi$, we can write explicitly the value of $P(X,t\vert\pi,0)$:
\begin{align}
P(X,t\vert\pi,0)=\frac{e^{-\frac{(X-\langle b\vert \mu^{\pi} \rangle)^2}{2\langle b\vert \gamma^{\pi} \vert b\rangle}}}{\left[2\pi \langle b\vert \gamma^{\pi} \vert b\rangle\right]^{1/2}}\label{PropagatorStartingFromPi}
\end{align}
Then, the average $\mu_i^{\pi,X}$ and covariance $\gamma_{ij}^{\pi,*}$ of the modes $a_i$ at $t$ given that one observed $X$ at $t$ and that the initial distribution is $\pi$ are obtained by the projection formulas (\ref{ProjectionMean},\ref{ProjectionCovariance}):
\begin{align}
\mu_i^{\pi,X}=\mu_i^{\pi}-\frac{\langle  e_i\vert\gamma^{\pi}\vert b \rangle }{\langle  b\vert\gamma^{\pi}\vert b \rangle }(\langle b\vert \mu^{\pi} \rangle -X)  \ ; \
\gamma_{ij}^{\pi,*}=\gamma_{ij}^{\pi}-\frac{\langle  e_i\vert\gamma^{\pi}\vert b \rangle \langle b \vert\gamma^{\pi}\vert e_j \rangle }{\langle  b\vert\gamma^{\pi}\vert b \rangle }\label{ProjectionPi}
\end{align}

For the moment, we choose gaussian initial conditions $P_{\text{ini}}$, which has means $m_i^{\text{ini}}$ and covariance matrix $\sigma_{ij}^{\text{ini}}$.  
The moments $m_i^{\text{ini}}$ and $\sigma_{ij}^{\text{ini}}$ can be chosen so that the initial distribution is the stationary distribution restricted to configurations such that $X=X_0$. In the case of a non-diffusive variable, this choice of initial conditions is obtained by applying the projection formulas (\ref{ProjectionMean},\ref{ProjectionCovariance}) to the moments of the stationary distribution (given by  $m_i^{\text{stat}}=0$ and $\sigma_{ij}^{\text{stat}}=\delta_{ij}/\lambda_{i}$):
\begin{align}
	m_{i}^{\text{ini}}=\frac{X_0 b_i}{\lambda_i L^2} \ ;Ê\ 
	\sigma_{ij}^{\text{ini}}= \sigma_{ij}^{\text{stat},*} =\frac{\delta_{ij}}{\lambda_i}-\frac{b_i b_j}{\lambda_i\lambda_j L^2}\label{CovarianceStationaryProjected}
\end{align}
where we have noted $L^2$ the stationary value of the root mean square of the observable $X$:
\begin{equation}
	L^2=\langle b \vert \sigma^{\text{stat}}\vert b\rangle = \sum_{q=2}^{N} b_q^2/\lambda_q 
\end{equation}
In the case of a diffusive variable, one has to take care of the mode with vanishing eigenvalue $\lambda_1=0$. In this case, the moments $m_i^{\text{ini}}$ and $\sigma_{ij}^{\text{ini}}$ are found by taking the limit of small $\lambda_1$ in Eq. (\ref{CovarianceStationaryProjected}), we find:
\begin{align}
	m_{i}^{\text{ini}}=\frac{\delta_{i1}X_0}{b_1} \ ; \ 
	\sigma_{ij}^{\text{ini}}	=\sigma_{ij}^{\text{stat},*}	=
	\begin{cases}
	\delta_{ij}/\lambda_i & \text{if}  \ i,j\ge2\\
	-b_j/(b_1\lambda_j) & \text{if} \  j\ge2,i=1 \\
	\sum_{q=2}^N b_q^2/(\lambda_q b_1^2) & \text{if} \  i=j=1\\
	\end{cases}	
	\label{CovarianceStationaryProjectedDiffusive}
\end{align}
The quantities  $\mu_i^{\text{ini},X}$ and $\gamma_{ij}^{\text{ini},*}$ are deduced from $m_i^{\text{ini}}$ and $\sigma_{ij}^{\text{ini}}$ by replacing the superscript ``$\pi$'' by ``$\text{ini}$'' in Eqs. (\ref{PropagationPi},\ref{ProjectionPi}). Let us now define the functions $\phi$ and $\psi$ such that: 
\begin{align}
	&\langle X(t)\rangle^{\text{ini}}=\langle  b\vert \mu^{\text{ini}}\rangle =X_0 \phi(t)\label{MeanEffectivePropagatorNonDiffusive}  \\  
	&\langle \Delta X^2(t)\rangle^{\text{ini}}=\langle  b\vert\gamma^{\text{ini}}\vert b \rangle =\psi(t) \label{VarianceEffectivePropagatorNonDiffusive}
\end{align} 
The function $\phi$ describes how the average of $X$ evolves with the time when the initial condition is stationary with a value $X=X_0$, whereas $\psi(t)$ is the variance of $X$ (with the same initial conditions). Using the projection and propagation formulas, we can find the values of $\phi$ and $\psi$. In the case of a non-diffusive variable, we obtain:
\begin{align}
\phi(t)=\sum_{i=2}^{N}\frac{ b_i^2 e^{-\lambda_i t}}{\lambda_iL^2} \ ; \ \psi(t)=L^2 [1-\phi^2(t)] \  (\text{Non-diffusive variable})\label{PsiPhiNonDiffusive}
\end{align}
whereas in the case of a diffusive variable, we have:
\begin{align}
\phi(t)=1 \ ; \ \psi(t) =2b_1^2 t +2 \sum_{j\ge2}b_j^2(1-e^{-\lambda_j t})/\lambda_j \ (\text{Diffusive variable})\label{PsiPhiDiffusive}
\end{align}
The equations (\ref{PsiPhiNonDiffusive},\ref{PsiPhiDiffusive}) mean that the average value of $X$ remains constant with time if $X$ is diffusive, but decreases to a stationary value when $X$ is not diffusive. At long times, $\psi\sim L^2$ for a non-diffusive variable (that relaxes to a stationary value), whereas $\psi\sim b_1 t$ for a diffusive variable (that diffuses at long times). Let us finally write the explicit expression for $P(X,t\vert\text{ini},0)$:
\begin{align}
	P(X,t\vert\text{ini},0)	= \frac{e^{-\frac{(X-X_0\phi)^2}{2\psi}}}{(2\pi\psi)^{1/2}}\label{EffectivePropagator}
\end{align}
All the terms appearing in the self-consistent equations (5,6,7) of the non Markovian theory are explicitly written in this section in Eqs. (\ref{PropagatorStartingFromPi},\ref{ProjectionPi},\ref{CovarianceStationaryProjected},\ref{CovarianceStationaryProjectedDiffusive},\ref{EffectivePropagator}). Note that the means $m_i^{\pi}$ associated to the positions in the main text are easily calculated from the means associated to the modes $a_i$ derived in this Supplementary Information (that are abusively represented with the same notation $m_i^{\pi}$) by using the rotation matrix $Q$ defined in Eq. (\ref{DefinitionMatrixQ}). 

\subsection{Self-consistent equations for the non-markovian theory in dimension $d=3$}
\label{SectionGeneralisationDim3}
Here, we briefly describe  how the theory can be extended to the 3-dimensional case. Let us assume that the initial conditions are isotropic. Then, the reaction can take place anywhere on the target, and we can introduce the probability $\pi_{\Omega}(\vert \ve[a]\rangle)$ of reacting with a configuration $\vert \ve[a]\rangle$ given that the observable has angular coordinates $\Omega=(\theta,\phi)$ when the reaction takes place.
Let us introduce the usual basis of unit vectors for the spherical coordinates $(\ve[u]_r,\ve[u]_{\theta},\ve[u]_{\phi})$, and the coordinates $(a_{i,r},a_{i,\theta},a_{i,\phi})$ of $\ve[a]_i$ in this basis: $\ve[a]_i= a_{i,r}\ve[u]_r+a_{i,\theta}\ve[u]_{\theta}+a_{i,\phi}\ve[u]_{\phi}$. Due to the symmetry, we can separate the coordinates in the splitting distribution: 
\begin{equation}
	\pi_{\Omega}(\vert \ve[a]\rangle)=\pi_r(\vert a_r\rangle)\pi_{\theta}(\vert a_{\theta}\rangle)\pi_{\phi}(\vert a_{\phi}\rangle)\label{SeparationCoordinates}
\end{equation}
The gaussian approximation is written for each coordinate: we assume that the radial splitting distribution $\pi_r$ is a multivariate gaussian, with mean vector $m_{i}^{\pi,\parallel}$ and a covariance matrix $\sigma_{ij}^{\pi,\parallel}$, whereas the two perpendicular splitting distributions $\pi_{\theta}$ and $\pi_{\phi}$ are multivariate gaussian, with vanishing mean vector ($m_{i}^{\pi,\perp}=0$), and a covariance matrix $\sigma_{ij}^{\pi,\perp}$. The resulting self-consistent equations are very cumbersome, that is why in this paper we make the simplifying assumption of isotropy of the covariance matrix: we assume that $\sigma^{\pi,\parallel}=\sigma^{\pi,\perp}$, and we note the common value of these matrices $\sigma^{\pi}$.  We note $\mu_{i}^{\pi}$ and $\gamma_{ij}^{\pi}$ the radial mean vector and the covariance matrix of the distribution $P(\vert \ve[a]\rangle , t \vert \pi_{\Omega},0)$. 
Let us specifically assume that the initial distribution is the stationary distribution restricted to configurations such that $\|\ve[R]\|=R_0$. 
Let $P(\vert \ve[a]\rangle,t \vert P_{\text{ini},\Omega},0)$ be the distribution of modes at $t$ starting from an initial stationary distribution where the observable value is $\ve[R]_0=R_0\ve[u]_r(\theta,\phi)$ and has the angular spherical coordinates $\Omega=(\theta,\phi)$. As in Eq. (\ref{SeparationCoordinates}), we can  separate this distribution into $P_r P_{\theta} P_{\phi}$, where each function is a gaussian. We now rewrite the equation (4) in the main text in the space of modes and with an explicit average over the angles:
\begin{align}
\tau &P_{\text{stat}}(\vert \ve[a]\rangle)=\int_0^{\infty}dt \int d\Omega \left[P(\vert \ve[a]\rangle,t\vert \pi_{\Omega},0)-P(\vert \ve[a]\rangle,t\vert P_{\text{ini},\Omega},0)\right]	\label{EqStartDim3}
\end{align}
where $d\Omega=\sin\theta d\theta d\phi/(4\pi)$. 
We multiply both members by $\delta(\langle c\vert \ve[a]\rangle -\ve[R]_\text{f})$ and reinterpret this equation:
\begin{align}
	\tau P_{\text{stat}}(\ve[R]_{\text{f}})P_{\text{stat}}(\vert \ve[a]\rangle\vert \ve[R]_{\text{f}})=
		\int_0^{\infty}dt \int d\Omega [ P(\ve[R]_{\text{f}},t\vert \pi_{\Omega},0)P(\vert \ve[a]\rangle,t\vert \ve[R]_{\text{f}},t;\pi_{\Omega},0)
-P(\ve[R]_{\text{f}},t\vert \text{ini},0)P(\vert \ve[a]\rangle,t\vert \ve[R]_{\text{f}},t;P_{\text{ini},\Omega},0)]		\label{EquationReinterpretedDim3}
\end{align}
This equation is the equivalent in 3 dimensions of Eq. (\ref{EqIntegraleDim1Reinterpreted}). Integrating it over all the modes gives an estimation of the mean first reaction time:
\begin{align}
\tau P_{\text{stat}}(\ve[R]_{\text{f}})=
\int_0^{\infty}dt\int d\Omega [P(\ve[R]_{\text{f}},t \vert \pi_{\Omega},0)-P(\ve[R]_{\text{f}},t \vert P_{\text{ini},\Omega},0)]\label{EstimationReactionTimeDim3}
\end{align}
For symmetry reasons, we can assume that $\ve[R]_{\text{f}}=Z_{\text{f}}\ve[u]_z$ without loss of generality. The distribution $P(\ve[R]_{\text{f}},t \vert \pi_{\Omega},0)$ can be calculated by noting that for a given value of $\theta$, the radial component of $\ve[R]_{\text{f}}$ is $Z_{\text{f}}\cos\theta$ whereas its component in the $\theta$ direction is $-Z_{\text{f}}\sin\theta$:
\begin{align}
	P(Z_{\text{f}}\ve[u]_z,t\vert \pi_{\Omega},0)
&=P(\cos\theta Z_{\text{f}} ,t \vert \pi_{r},0)P(-\sin\theta Z_{\text{f}} ,t \vert \pi_{\theta},0)P(0,t \vert \pi_{\phi},0)\\
	&=\frac{1}{[2\pi\langle  b\vert\gamma^{\pi}\vert b \rangle]^{3/2}}\ e^{-\frac{1}{2}\left[\frac{(Z_{\text{f}}  \text{cos}\theta - \langle  b\vert\mu^{\pi}\rangle )^2}{\langle  b\vert\gamma^{\pi}\vert b\rangle }+\frac{(-Z_{\text{f}} \text{sin}\theta)^2}{\langle  b\vert\gamma^{\pi}\vert b \rangle }\right]}	
\end{align}
In the same way, we have:
\begin{align}
P&(Z_{\text{f}}\ve[u]_z,t\vert P_{\text{ini},\Omega},0)=\frac{e^{-\frac{1}{2\psi}\left[(Z_{\text{f}}  \text{cos}\theta - R_0\phi )^2+(Z_{\text{f}} \text{sin}\theta)^2\right]}	
}{[2\pi\psi]^{3/2}}
\end{align}
Obviously, the equation (\ref{EstimationReactionTimeDim3}) can be simplified by taking $Z_{\text{f}}=0$, in which case we obtain:
\begin{align}
\tau &P_{\text{stat}}(\ve[0])=\int_0^{\infty}dt [P(\ve[0],t \vert \pi_{\Omega},0)-P(\ve[0],t \vert P_{\text{ini},\Omega},0)]
\label{EquationTempsDim3RfNul}
\end{align}
with:
\begin{align}
&P(\ve[0],t\vert \pi_{\Omega},0)	=\frac{1}{[2\pi\langle  b\vert\gamma^{\pi}\vert b \rangle]^{3/2}}\ e^{-\frac{(\langle  b\vert\mu^{\pi}\rangle )^2}{2\langle  b\vert\gamma^{\pi}\vert b\rangle }}	 \\
&P(\ve[0],t\vert P_{\text{ini},\Omega},0)=\frac{1}{[2\pi\psi]^{3/2}}\ e^{-\frac{(R_0\phi )^2}{2\psi}}
\end{align}

The self-consistent equations that define the values of $m_{i}^{\pi}$ are obtained by multiplying Eq. (\ref{EquationReinterpretedDim3}) by $a_{iz}$ and by integrating over all the modes. To do so, we first need to evaluate the integral:
\begin{align}	
\int d\vert\ve[a]\rangle a_{iz}P(\vert \ve[a]\rangle ,t\vert Z_{\text{f}} \ve[u]_z,t ; \pi_{\Omega},0)
&=\int d\vert \ve[a]  \rangle (a_{ir} \text{cos}\theta-a_{i\theta}\text{sin}\theta)P(\vert \ve[a]\rangle ,t\vert Z_{\text{f}}\ve[u]_z,t ; \pi_{\Omega},0)\nonumber\\
&=\text{cos}\theta \mu_{i}^{\pi,0}+ Z_{\text{f}} \frac{\langle  e_i\vert\gamma^{\pi}\vert b \rangle }{\langle  c\vert\gamma^{\pi}\vert b \rangle} \label{Eq568}
\end{align}
In the passage from the first to the second line, we used the fact that $a_{iz}=a_{ir} \text{cos}\theta-a_{i\theta}\text{sin}\theta$. The passage to the third line uses two times the projection formula (\ref{ProjectionMean}).
In principle, we can choose any value for $Z_{\text{f}} \in[0;a]$. In order to obtain simple formulas, we chose to develop the equations for the first moment at lowest order in $Z_{\text{f}}$. Reporting Eq. (\ref{Eq568}) into the equation (\ref{EquationReinterpretedDim3}) (multiplied by $a_{iz}$) and integrating over all the modes leads at lowest order in $Z_{\text{f}}$ to:
\begin{align}
\tau &P_S(\ve[0]) K_i=\nonumber\\
&\int_0^{\infty}dt \Bigg[\left(\frac{1}{3}\mu_{i}^{\pi,0}\frac{\langle  b\vert\mu^{\pi}\rangle }{\langle  b\vert\gamma^{\pi}\vert b \rangle }+ \frac{\langle  e_i\vert\gamma^{\pi}\vert b \rangle }{\langle  b\vert\gamma^{\pi}\vert b \rangle }\right)P(\ve[0],t\vert\pi_{\Omega},0)
-\left(\frac{1}{3}\mu_{i}^{\text{ini},0}\frac{R_0\phi}{\psi}+ \frac{\langle  e_i\vert\gamma^{\text{ini}}\vert b \rangle }{\psi}\right)P(\ve[0],t\vert P_{\text{ini},\Omega},0)
\Bigg]\label{EquationFirstMomentDim3}
\end{align}
where 
\begin{align}
	K_i=
	\begin{cases}
	\delta_{i1}/b_1 & \text{(diffusive variable)}\\
	b_i/(\lambda_i L^2) & \text{(non-diffusive variable)}
	\end{cases}
\end{align}
We do not give any details on the derivation of the equations that define the second moments at the limit $Z_{\text{f}}=0$, we multiply Eq. (\ref{EquationReinterpretedDim3}) by $a_{iz}a_{jz}$ and integrate over the modes, we obtain: 
\begin{align}
\tau P_{\text{stat}} (\ve[0])&\sigma_{ij}^{\text{stat},*}=
\int_0^{\infty}dt
\Bigg[\left(\frac{\mu_{i}^{\pi,0}\mu_{j}^{\pi,0}}{3}+\gamma_{ij}^{\pi,*}\right)
P(\ve[0],t\vert\pi_{\Omega},0)
-\left(\frac{\mu_{i}^{\text{ini},0}\mu_{j}^{\text{ini},0}}{3}+\gamma_{ij}^{\text{ini},*}\right)P(\ve[0],t\vert P_{\text{ini},\Omega},0)\Bigg]\label{EquationSecondMomentDim3}
\end{align}
The equations (\ref{EquationTempsDim3RfNul}), (\ref{EquationFirstMomentDim3}) and (\ref{EquationSecondMomentDim3}) form a set of non-linear equations that enable to compute both the mean first reaction time and the moments of the splitting probability distribution under the hypothesis of isotropy of the covariance matrix. They are the generalization to a 3--dimensional space of the equations (5,6,7) of the main text.

\subsection{Mean first passage time averaged over initial conditions (non-diffusive observable)}
In the case of a non-diffusive variable, we can ask for the time it takes for the reaction when the initial distribution is the stationary distribution (restricted to configurations that lie outside the reactive region, $\|\ve[R] \| \ge a$). Then, the initial distribution is a superposition of the distributions restricted to a given value of $R_0$:
\begin{align}
	P_{\text{ini}}(\vert a\rangle)=P_{\text{stat}}(\vert a\rangle\vert \|\ve[R] \| \ge a)=\int_a^{\infty}dR_0 R_0^2 P_{\text{stat}}(R_0 ) P_{\text{stat}}(\vert a\rangle\vert \|\ve[R]\| = R_0 ) \label{DefinitionAverageOverInitialCondition}
\end{align}
with:
\begin{align}
P_{\text{stat}}(R_0 )=\frac{e^{-R_0^2/(2L^2)}}{Z(a,L^2)} \ ; \
Z(a,h)=\int_a^{\infty}dR_0 R_0^2 e^{-R_0^2/(2h)} \label{DefinitionZ}
\end{align}
Eq. (\ref{DefinitionAverageOverInitialCondition}) states that the initial distribution is a superposition of the initial distributions studied in the section \ref{SectionGeneralisationDim3}. We can therefore follow step by step the calculations that led to the equations (\ref{EquationTempsDim3RfNul}), (\ref{EquationFirstMomentDim3}) and (\ref{EquationSecondMomentDim3}) and average over $R_0$ at the end of the calculation. For simplicity, we only give the self-consistent equations for the first moments of the splitting when the covariance matrix $\sigma_{ij}^{\pi}$ is approximated by $\sigma_{ij}^{\text{stat},*}$. This approximation turns out to be an excellent approximation. Under this approximation, the equation that defines the mean vector of the splitting is: 
\begin{align}
\int_0^{\infty}dt \Bigg\{&\left(\frac{1}{3}\mu_{i}^{\pi,0}\frac{\langle  b\vert\mu^{\pi}\rangle }{\psi}+ \frac{\langle  e_i\vert\gamma^{\{X_0,\text{ini}\}}\vert b \rangle }{\psi}-\frac{b_i}{\lambda_i L^2}\right)P(\ve[0],t\vert\pi_{\Omega},0)\nonumber\\
&-\left[\frac{\phi}{3\psi}\left(\frac{b_ie^{-\lambda_i t}}{\lambda_i L^2}-\phi\frac{\langle e_i\vert \gamma^{\{X_0,\text{ini}\}}\vert b\rangle}{\psi}\right)\frac{G(a,\psi)}{Z(a,L^2)}+\left(\frac{\langle e_i\vert \gamma^{\{X_0,\text{ini}\}}\vert b\rangle}{\psi}-\frac{b_i}{\lambda_i L^2}\right)\frac{Z(a,\psi)}{Z(a,L^2)} \right]\frac{1}{(2\pi\psi)^{3/2}}\Bigg\} =0 \label{EqFirstMomentGlobal}
\end{align}
where the superscript ``$\{R_0,\text{ini}\}$'' simply indicates that one refers to the only initial conformations such that $\|\ve[R] \| =R_0$, the function $Z$ is defined in Eq. (\ref{DefinitionZ}) and $G$ is defined by:
\begin{align}
G(a,h)=\int_a^{\infty}dR_0 R_0^4 e^{-R_0^2/(2h)} = a h (a^2+3 h) e^{-a^2/(2h)}+3\sqrt{\frac{\pi}{2}}h^{5/2}\left[1-\text{Erf}\left(a/\sqrt{2h}\right)\right]
\end{align}
The mean first passage time can be evaluated by using the formula:
\begin{align}
\tau P_{\text{stat}}(\ve[R]_{\text{f}})=
\int_0^{\infty}dt \int d\Omega \left[P(\ve[R]_{\text{f}},t \vert \pi_{\Omega},0)-\int_a^{\infty}dR_0 R_0^2 P_{\text{stat}}(R_0)P(\ve[R]_{\text{f}},t\vert \{R_0,\text{ini}\},\Omega,t=0) \right]\label{tauGlobalRfFini}
\end{align}
or, taking $\ve[R]_{\text{f}}=\ve[0]$: 
\begin{align}
\tau P_{\text{stat}}(\ve[0])=
\int_0^{\infty}dt \left[P(\ve[0],t \vert \pi_{\Omega},0)-\frac{Z(a,\psi)}{Z(a,L^2)(2\pi\psi)^{3/2}}\right]\label{tauGlobal0}
\end{align}
We note that, in the limit $a\ll L$, we have $Z(a,\psi)/Z(a,L^2)\simeq\psi^{3/2}/L^3$ and we obtain the simpler expression from Eq. (\ref{tauGlobalRfFini}):
\begin{align}
\tau P_{\text{stat}}(\ve[R]_{\text{f}})=
\int_0^{\infty}dt \left[P(\ve[R]_{\text{f}},t \vert \pi_{\Omega},0)-P_{\text{stat}}(\ve[R]_{\text{f}})\right]\label{tauGlobalSmallTarget}
\end{align}

This expression enables us to make the link between the Markovian theory and the Wilemski-Fixman theory with a delta-sink function. The Markovian approximation is written as: $\sigma_{ij}^{\pi}=\sigma_{ij}^{\text{stat},*}$, and $m_i^{\pi}=m_i^{\text{stat},a}$. Reporting these approximations into Eq. (\ref{tauGlobalSmallTarget}) and taking $Z_f=a$ leads to the expression of the Markovian approximation of $\tau$ in the case of a small target:
\begin{align}
	\tau\simeq	
\int_0^{\infty} dt \left\{\frac{L^2}{ a^2 \phi(1-\phi^2)^{1/2} }e^{\frac{a^2}{2L^2}}e^{-\frac{a^2+a^2\phi^2}{2L^2(1-\phi^2)}}\text{sinh}\left[\frac{a^2\phi}{L^2(1-\phi^2)}\right]
-1\right\}
\end{align}
This expression is equivalent to the Wilemski-Fixman theory with the delta-sink approximation introduced in Ref. \cite{Pastor1996}. This shows that the Markovian approximation leads to the same results as the Wilemski-Fixman theory.

\section{Scaling relations}
In this section, we derive scaling relations in 3 dimensions. We always derive scaling relations by assuming that the second moment of the splitting distribution can be approximated by its stationary value ($\sigma^{\pi}\simeq\sigma^{\text{stat},*}$). We first identify the relevant regimes by looking at the time scales.

\subsection{Properties of the walk at different time scales}
At short times, in both diffusive and non-diffusive cases, we have:
\begin{equation}
	\psi(t)\simeq 2 D_0 t \ ; \ D_0=\langle b\vert b\rangle \hspace{1cm}Ê (t\rightarrow0) \label{ComportementFaiblesTemps}
\end{equation}
Hence, the walk is always diffusive at short times with an effective diffusion coefficient $D_0$. In the case that the observable is the first monomer position, $D_0=D$, whereas $D_0=2D$ in the case of the end-to-end vector.
At long times, for a non-diffusive observable, we have:
\begin{align}
	\psi \simeq L^2 \ ; \ \phi \simeq b_2^2 e^{-\lambda_2 t}/(L^2\lambda_2)
\end{align} 
This clearly shows that the observable relaxes to a stationary distribution. In the case of a diffusive variable, we have:
\begin{align}
	\psi(t)=2 b_1^2 t 
\end{align}
In this case, the motion is diffusive at long times. In the case that the observable is the first monomer position, the diffusion coefficient at large times is simply the diffusion coefficient of the center of mass $b_1=D_{\text{CM}}=D/N$. 
Irregular behavior of $\psi$ and $\phi$ appears at the limit of infinite $N$. In this case, one sums over an infinite number of modes, and the eigenvalues given by Eq. (\ref{ValeurDesValeursPropres}) become in the large $N$ limit:
\begin{align}
	\lambda_q\simeq (q-1)^2/\tau_{\text{R}}\label{BehaviorSmallQ}
\end{align}
where $\tau_{\text{R}}=N^2/\pi^2$ is the Rouse time, the slowest relaxation time of the non-diffusive modes.
In the case of the position of the first monomer, $b_1=1/\sqrt{N}$ and $b_{i\ge2}\simeq\sqrt{2/N}$. 
Let us call $\psi_{\infty}$ and $\phi_{\infty}$ the functions $\psi$ and $\phi$ obtained in the limit of an infinite number of modes.
The short time limit of $\psi_{\infty}$ can be evaluated by transforming the (infinite) sums (\ref{PsiPhiNonDiffusive},\ref{PsiPhiDiffusive}) into integrals (see \cite{KhokhlovBook} for details):
\begin{align}
	\psi(t) \simeq \begin{cases}
	4\sqrt{t/\pi} Ê& \text{if } \ve[R]=\ve[r]_1\\
	8\sqrt{t/\pi} Ê& \text{if } \ve[R]=\ve[r]_N-\ve[r]_1
	\end{cases}
\end{align}
This behavior defines an effective walk dimension: $\psi\sim t^{2/d_w}$ with $d_w=4$.

\subsection{The limit of small target size $a$ (at fixed $N$)}
\label{SectionLimitSmallSize}
\subsubsection{Case of a non-diffusive variable}
Let us write Eq. (\ref{EqFirstMomentGlobal}) in the case that $\sigma^{\pi}\simeq\sigma^{\text{stat},*}$ and in the case that $a\ll L$. 
\begin{align}
&\int_0^{\infty}dt \left[\frac{\Gamma}{3\psi}\left(m_{i}^{\pi} e^{-\lambda_i t}-\frac{\Gamma b_i (1-e^{-\lambda_i t}\phi)}{\psi \lambda_i}\right)+\frac{b_i}{\lambda_i}\left(\frac{1-e^{-\lambda_i t}\phi}{\psi}-\frac{1}{L^2}\right)
\right]\frac{e^{-\frac{\Gamma^2}{2\psi}}}{(2\pi \psi)^{3/2}}=0 \label{EG7690}
\end{align}
where we have set $\Gamma=\langle  b\vert\mu^{\pi}\rangle$ and we have used the fact that: $\gamma^{\{\text{stat},*\}}=\gamma^{\{X_0,\text{ini}\}}$ and that:
$\langle e_i \vert \gamma^{\{\text{stat},*\}} \vert b \rangle  =b_i /\lambda_i (1- e^{-\lambda_i t} \phi)$.
We assume the scaling $m_{i}^{\pi}=a\ \tilde{m}_{i}$ ; $\Gamma(t)=a\ \tilde{\Gamma}(t)$ for $a\rightarrow0$. We check this scaling by calculating all the terms of Eq. (\ref{EG7690}) in the limit $a\rightarrow0$ by taking their short time expression. For example, we have:
\begin{align}
&\int_0^{\infty}dt \frac{\Gamma m_{i,r}^{\pi} e^{-\lambda_i t}}{3\psi}\frac{e^{-\frac{\Gamma^2}{2\psi}}}{(2\pi \psi)^{3/2}}\simeq \int_0^{\infty}dt \frac{a^2 \ \tilde{m}_{i} }{3(2D_0t)}\frac{e^{-a^2/[2(2D_0t)]}}{[2\pi(2D_0t)]^{3/2}}=\frac{\tilde{m}_i}{12 \pi D_0 a}
\end{align}
All the other terms of Eq. (\ref{EG7690}) are also of order $O(a^{-1})$, which proves that the development for small $a$ is consistent. Equating all the coefficients of $a^{-1}$ gives the following estimation for the moments:
\begin{equation}
	m_{i}^{\pi}\simeq a\ \tilde{m}_i=a \frac{b_i}{\lambda_i}\left(-\frac{\lambda_i}{D_0}+\frac{1}{L^2}\right)
\end{equation}
Note that this formula is consistent with the condition $\langle b\vert m^{\pi}\rangle=a$ (because $\langle b\vert b\rangle=D_0$). The conclusion is that the moments of the splitting probability vanish for small target size, and therefore they do not play any role in the global mean reaction time: the Markovian theory gives the same result as the non Markovian theory in this limit. This time can be evaluated by replacing $m_i^{\pi}$ by 0 in Eq. (\ref{tauGlobal0}) and then evaluating for small $a$ the resulting integral:
\begin{align}
	\tau P_{\text{stat}}(\ve[0])\simeq \int_0^{\infty} dt\  \frac{e^{-a^2/[2(2D_0t)]}}{[2\pi(2D_0t)]^{3/2}}=\frac{1}{4\pi D_0 a}
\end{align}
This behavior of $\tau$ is the same in the Markovian theory and the non Markovian theory and is fully compatible with the results of the other classical Markovian theory (the harmonic spring model \cite{SUNAGAWA1975,Szabo1980}). The monomers behave as if they were disconnected in an effective volume $V_{\text{eff}}=1/P_{\text{stat}}(\ve[0])$. 

\subsubsection{Case of a diffusive variable}
We now derive the behavior of $\tau$ and $m_{i}^{\pi}$ in the case of a diffusive variable in the limit $a\rightarrow0$ at fixed initial distance $R_0$. We write Eq. (\ref{EquationFirstMomentDim3}) for $i\ge2$ for a large initial distance $R_0\rightarrow\infty$:
\begin{align}
&\int_0^{\infty}dt \left[\frac{\Gamma}{3\psi}\left(m_{i}^{\pi} e^{-\lambda_i t}-\frac{\Gamma b_i (1-e^{-\lambda_i t})}{\psi \lambda_i} \right)+\frac{ b_i (1-e^{-\lambda_i t})}{\psi \lambda_i}\right]\frac{e^{-\frac{\Gamma^2}{2\psi}}}{(2\pi \psi)^{3/2}}=0
\end{align}
where we have set $\Gamma=\langle  b\vert\mu^{\pi}\rangle$. We apply again the method that we described in the case of a non-diffusive observable. We assume the scaling $m_{i}^{\pi}=a\ \tilde{m}_{i},\ \Gamma(t)=a\ \tilde{\Gamma}(t)$ for $a\rightarrow0$. This scaling is consistent and leads to the following evaluation of the moments:
\begin{align}
	m_{i}^{\pi}=
	\begin{cases}
		-a b_i/D_0 & \text{if } i\ge2\\
		a ( 2/b_1-b_1/D_0) & \text{if } i=1
	\end{cases}
\end{align}
where the second equality is deduced from the condition $\langle b\vert m^{\pi}\rangle=a$. Note that, in the case that the observable is the first monomer, this equation implies that the average position of the center-of-mass at the reaction is:
\begin{align}
	\langle x_{\text{cm}}\rangle_{\pi}=b_1 m_{1}^{\pi}=a\left(1+\frac{\sum_{i=2}^{N}b_i^2 }{b_1^2+\sum_{i=2}^{N}b_i^2 }\right)\ge a
\end{align}
That is to say, the position of the center of mass at the reaction is at the exterior of the target, at a distance of order $a$, as expected from the intuition. In this case also, the average reaction time is evaluated to be:
\begin{align}
	\tau \simeq  V \int_0^{\infty} dt\  \frac{e^{-a^2/[2(2D_0t)]}}{[2\pi(2D_0t)]^{3/2}}=\frac{V}{4\pi D_0 a}
\end{align}
This last equation shows that, in this regime, the first monomer behaves as if it were not connected to the rest of the polymer chain.

\subsection{Scaling in the thermodynamic limit $N\rightarrow\infty$}
\subsubsection{Reaction between the first monomer and a target.}

\begin{figure}[h!]
 \includegraphics[width=8cm,clip]{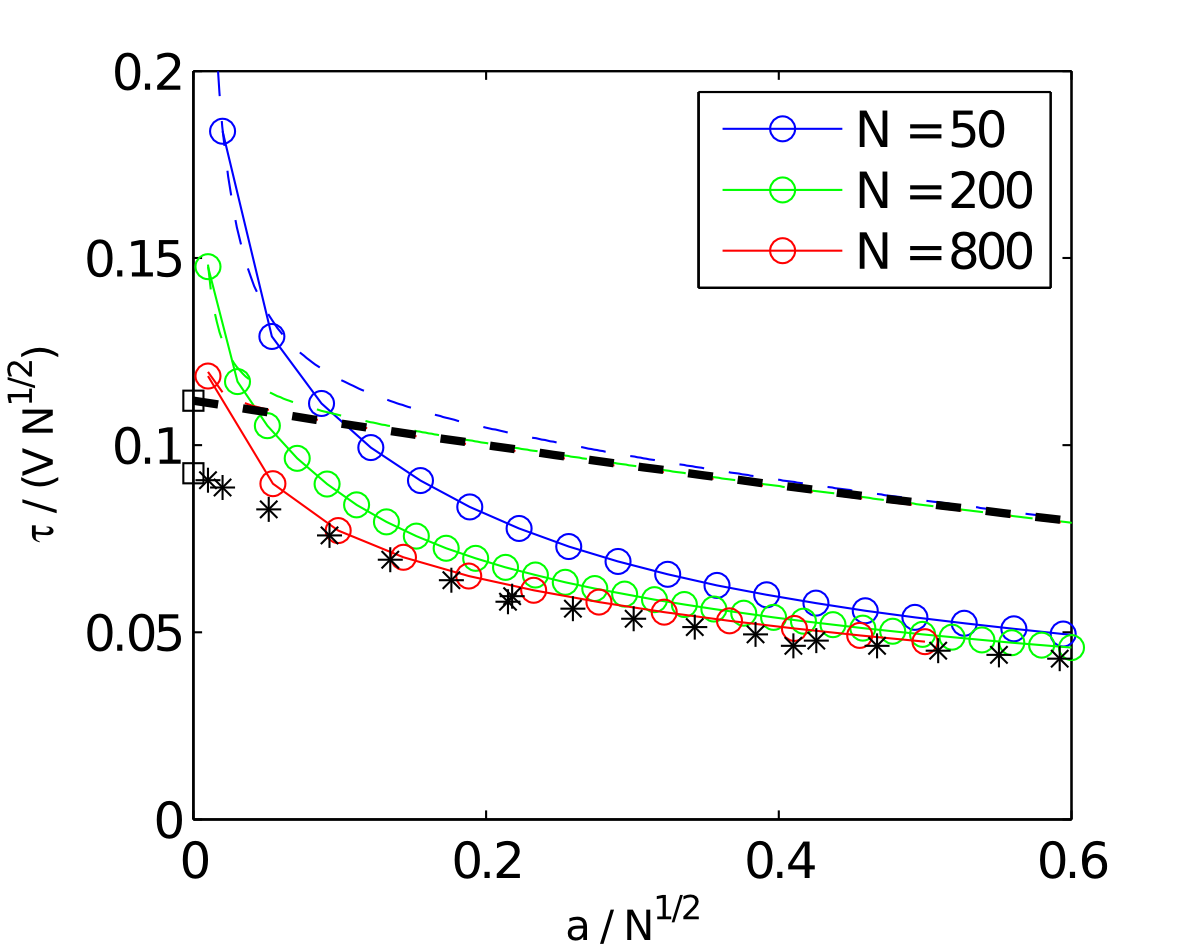}   
 \caption{{\bf Determination of the scaling relation for the intermolecular reaction time averaged over initial distances.} Circles: solution of Eq. (\ref{EquationFirstMomentDim3}) valid for finite $N$ for various values of $N$ (represented with a color code). The dashed lines represent the prediction of the Markovian theory for the same values of parameters. The black stars represent the solution of Eq. (\ref{EqMomentRescaleNonDiff}) when truncated at a  large enough  $N_c$ and give an estimation of the function $f$ [Eq. (\ref{tauDiffLargeN})], while $f_{\text{Markovian}}$  is represented by the black dashed line.
\label{FigSIDiffusif}}
\end{figure}

Here, we determine the scaling relations for the mean first passage time at the thermodynamic limit $N\rightarrow\infty$ in the case that the observable is $\ve[R]=\ve[r]_1$. In the limit of large $N$, the number of modes that must be taken account is infinite, the Rouse eigenvalues are approximated by $\lambda_q\simeq (q-1)^2\pi^2/N^2$, and the coefficients $b_q$ are approximated by $b_{q\ge2}\simeq\sqrt{2/N}$. We introduce the rescaled variables $\tilde{m}_q=m_q/N$, $\tilde{a}=a/\sqrt{N}$, $\tilde{\tau}=\tau/N^2$, $\tilde{b}_q=b_q\sqrt{N}$, $\tilde{V}=V/N^{3/2}$. In terms of these rescaled variables, one obtains the equations valid for $q\ge1$:
\begin{align}
&\int_0^{\infty}dt \left[\frac{\tilde{\Gamma}}{3\psi_{\infty}}\left(\tilde{m}_{q+1} e^{-\pi^2 q^2 t}-\frac{\tilde{\Gamma} \sqrt{2} (1-e^{-q^2\pi^2 t})}{\psi_{\infty}q^2\pi^2} \right)+\frac{ \sqrt{2} (1-e^{-q^2\pi^2 t})}{\psi_{\infty} q^2\pi^2}\right]\frac{e^{-\frac{\tilde{\Gamma}^2}{2\psi_{\infty}}}}{(2\pi \psi_{\infty})^{3/2}}=0	\label{EqMomentRescaleDiff}
\end{align}
the first moment $\tilde{m}_1$ is determined such that $\tilde{m}_1+\sqrt{2}\sum_{q\ge2}\tilde{m}_q=\tilde{a}$, which leads to:
\begin{align}
	\tilde{\Gamma}=\tilde{a}+\sum_{q=1}^{\infty}\tilde{m}_{q+1}^{\pi}(e^{-\pi^2q^2 t}-1) \ ; \ \psi_{\infty}=2 t + 4 \sum_{q=1}^{\infty} \frac{1-e^{-q^2\pi^2 t}}{q^2\pi^2}
\end{align}
and the evaluation of the mean first passage time is :
\begin{align}
	\tilde{\tau} =\tilde{V} \int_0^{\infty}dt \frac{e^{-\frac{\tilde{\Gamma}^2}{2\psi_{\infty}}}}{(2\pi \psi_{\infty})^{3/2}}=\tilde{V}f(\tilde{a})\label{tauDiffLargeN}
\end{align}
This defines a scaling relation $\tau/V=\sqrt{N} f(\tilde{a})=\sqrt{N}f(a/\sqrt{N})$, where $f$ is a dimensionless function that can be evaluated numerically.
The first numerical approach consists in trying to solve the exact equations (\ref{EquationFirstMomentDim3}) for $N$ large. This method gives the exact value of the mean first reaction time for any finite $N$, but is not suitable to determine the mean first reaction time in the limit of a small target. 
Indeed, it can be seen that all the curves obtained with this method converge when $N\rightarrow\infty$ to a single curve $f(\tilde{a})$ (Fig \ref{FigSIDiffusif}), but that the convergence is very slow for small target sizes. This method is therefore not suitable to estimate $f(\tilde{a})$ for $\tilde{a}<0.2$. The reason of this difficulty comes from the finite size effect detailed in section \ref{SectionLimitSmallSize}: in the limit of small size $\tilde{a}\ll1/N$, the mean first passage time scales as $1/\tilde{a}$. To overcome this difficulty, we directly approach the solution of the rescaled equations (\ref{EqMomentRescaleDiff}) by introducing a cutoff $N_c$ beyond which we approximate the moments by 0 ($m_q=0$ if $q\ge N_c$). The solutions obtained for finite $N_c$ are expected to converge for $N_c\rightarrow\infty$, leading to the determination of $f(\tilde{a})$ without having the problem of the finite size effect, as can be seen on Fig \ref{FigSIDiffusif}. This method enables us to determine the whole function $f(\tilde{a})$.

The scaling in the Markovian approach is much more straightforward. Inserting $\tilde{\Gamma}\simeq \tilde{a}$ into Eq. (\ref{tauDiffLargeN}) yields:
\begin{align}
	\tilde{\tau}_{\text{Markovian}} =\tilde{V} \int_0^{\infty}dt \frac{e^{-\frac{\tilde{a}^2}{2\psi_{\infty}}}}{(2\pi \psi_{\infty})^{3/2}}=\tilde{V}f_{\text{Markovian}}(\tilde{a})
\end{align}
Therefore, the Markovian estimate for $f(0)$ is $f(0)=\int dt (2\pi \psi_{\infty})^{-3/2}\simeq 0.112$, which gives the estimate $a_{\text{eff}}/\sqrt{N}=1/[4\pi f(0)]=0.71$. The difference with the non Markovian result is approximately $21\%$, and reaches $100\%$ for intermediate target sizes. The effective target size is related to the function $f$ by:
\begin{align}
	a_{\text{eff}}=\frac{l_0\sqrt{N}}{4\pi f(\tilde{a})}
\end{align}

\subsubsection{Cyclization reaction}
\begin{figure}[h!]
 \includegraphics[width=8cm,clip]{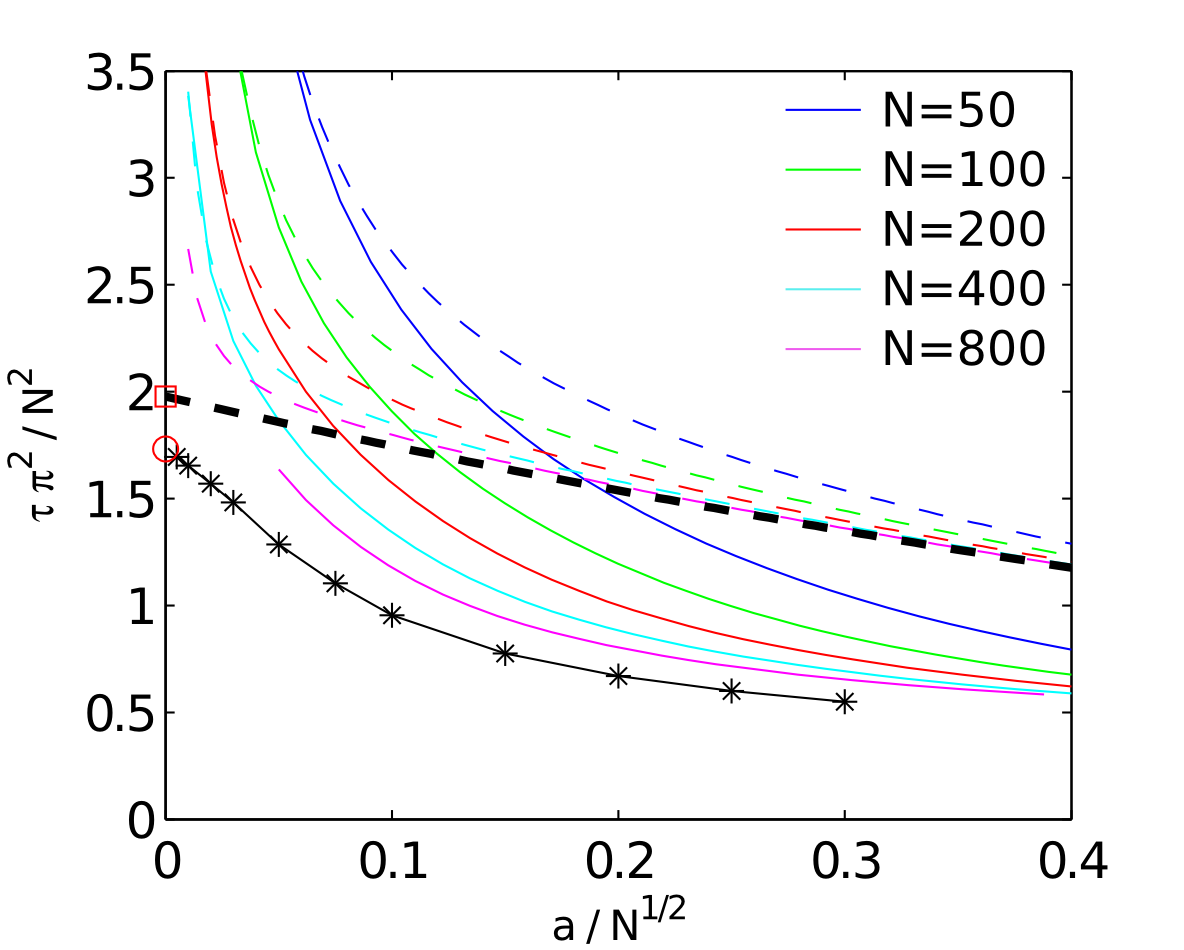}   
 \caption{{\bf Determination of the scaling relation for time the cyclization time averaged over stationary initial conditions.} 
Continuous lines: global mean cyclization time predicted by the non Markovian theory, obtained by solving Eq. (\ref{EquationFirstMomentDim3})  for various values of $N$ (represented with a color code). The dashed lines represent the prediction of the Markovian theory for the same values of $N$. The black stars represent the solution of Eq. (\ref{EqMomentRescaleNonDiff}) when truncated at a large enough $N_c$, and represent the function $c$ appearing in Eq. (\ref{tauNonDiffLargeN}). The black dashed line represents $c_{\text{Markovian}}$ [Eq. (\ref{tauDiffLargeNMarkovian})].}
\label{FigSICyclization}
\end{figure}

In the case of the cyclization reaction, we apply the same method. The rescaled equations read, for $q$ odd:
\begin{align}
&\int_0^{\infty}dt \left[\frac{\tilde{\Gamma}}{3\psi_{\infty}}\left(\tilde{m}_{q} e^{-\pi^2 q^2 t}-\frac{\tilde{\Gamma} 2\sqrt{2} (1-e^{-q^2\pi^2 t}\phi_{\infty})}{\psi_{\infty}q^2\pi^2} \right)+\frac{ 2\sqrt{2}}{q^2\pi^2} \left(\frac{1-e^{-q^2\pi^2 t}\phi_{\infty}}{\psi_{\infty} }-1\right)\right]\frac{e^{-\frac{\tilde{\Gamma}^2}{2\psi_{\infty}}}}{(2\pi \psi_{\infty})^{3/2}}=0	\label{EqMomentRescaleNonDiff}
\end{align}
with:
\begin{align}
	\tilde{\Gamma}=\sum_{q=1,\text{odd}}^{\infty}\tilde{m}_{q}^{\pi}e^{-\pi^2q^2 t} \ ; \ \phi_{\infty}= \sum_{q=1,\text{odd}}^{\infty} \frac{8 \ e^{-q^2\pi^2 t}}{q^2\pi^2} \ ; \ \psi_{\infty}=1-\phi_{\infty}^2
\end{align}
and the evaluation of the mean first passage time is :
\begin{align}
	\tilde{\tau} =\frac{\tau}{N^2} = (2\pi)^{3/2} \int_0^{\infty}dt \left(\frac{e^{-\frac{\tilde{\Gamma}^2}{2\psi_{\infty}}}}{(2\pi \psi_{\infty})^{3/2}}-\frac{1}{(2\pi)^{3/2}}\right)=\frac{c(\tilde{a})}{\pi^2}\label{tauNonDiffLargeN}
\end{align}
We apply the same method as in the case of the intermolecular reaction to determine the dimensionless function $c$. One additional difficulty in this case is that the truncated equations do not allow a solution such the solution such that $\sum_{q=1}^{N_c}m_q=\tilde{a}$ because the redundancy of the equations is lost when the equations are truncated. We have therefore to release one equation and to impose the condition  $\sum_{q=1}^{N_c}m_q=\tilde{a}$. We tried to leave the restriction on the first mode as well as the last one, leading to almost no difference in the limit of large $N$.  The results are presented on Fig. \ref{FigSICyclization}.

We find the scaling $\tau\simeq 1.732 \tau_R$ in the limit of small target size. The Markovian estimation gives:
\begin{align}
	\tilde{\tau}_{\text{Markovian}} =\frac{\tau}{N^2} = (2\pi)^{3/2} \int_0^{\infty}dt \left(\frac{e^{-\frac{(\tilde{a}\phi_{\infty})^2}{2\psi_{\infty}}}}{(2\pi \psi_{\infty})^{3/2}}-\frac{1}{(2\pi)^{3/2}}\right)=\frac{c_{\text{Markovian}}(\tilde{a})}{\pi^2}\label{tauDiffLargeNMarkovian}
\end{align}
From Eq. (\ref{tauDiffLargeNMarkovian}), we estimate the Markovian scaling relation for very small target sizes to be $\tau\simeq 1.977\tau_R$ for small target sizes, the difference with the non Markovian scaling is about $14\%$. For realistic target sizes (for example for $\tilde{a}=0.2$), $c\simeq0.5c_{\text{Markovian}}$: the difference between the Markovian and non Markovian scaling is of the order of $100\%$.

\section{Details of simulations}

\subsection{Simulation of cyclization reaction}

We performed simulations of cyclization events by using the algorithm presented in Ref. \cite{Pastor1996}. The initial condition is an equilibrium configuration of the polymer, in which the coordinates of each mode $a_{i,x},a_{i,y},a_{i,z}$ can be chosen from a normal distribution with variance $1/\lambda_i$ and zero mean. The initial positions of the monomers are deduced from the initial values of the modes by using the transformation rule of Eq. (\ref{DefinitionModes}). If the initial configuration is inside the reactive zone ($\vert \ve[r]_{\text{ee}}\vert< a$), this configuration is rejected and another configuration is chosen. After the determination of the initial configuration, the  monomer positions evolves at each time step with the following algorithm:
\begin{equation}
	x_i(t_{k+1})=	x_i(t_{k}) -\left[\sum_{j=1}^N M_{ij} x_j(t_k)\right] (\Delta t)_k +  \sqrt{2\ (\Delta t)_k}\ u_{k,x}
\end{equation}
where $u_{k,x}$ is a number generated with the normal distribution with variance unity. The evolution of the other coordinates follows the same equation, and the time step is variable \cite{Pastor1996}:
\begin{equation}
(\Delta t)_k=t_{k+1}-t_{k}=
\begin{cases}
(\Delta t)_{\text{low}}+(\Delta t)_{\text{high}} \ \text{sin}\left(\left\vert \ve[R]_{ee}(t_k)\vert ^2-a^2\right)\pi/6\right) &  \text{if } \vert \ve[R]_{ee}(t_k)\vert^2 < a^2+3\\
(\Delta t)_{\text{low}}+(\Delta t)_{\text{high}} & \text{otherwise}
\end{cases}
\end{equation}
With this choice, the time steps become smaller and smaller when the reactive zone is approached. The slight differences of numerical constants with Ref.  \cite{Pastor1996} is due to a slightly different choice of units. The simulation runs until the condition $\vert \ve[R]_{ee}\vert<a$ is satisfied, in which case the value of $\tau=\sum_k (\Delta t)_k$ is an estimation of the first cyclization time for this trajectory. We have chosen $(\Delta t)_{\text{high}}=10^{-4}$, and $(\Delta t)_{\text{low}}=10^{-7}$. These values are approximately 3 times smaller than the corresponding values in Ref. \cite{Pastor1996}.

\subsection{Simulation of intermolecular reactions}
We now describe the simulations that lead to the estimation of the mean first passage time of the first monomer to a target located at $x=0$, when there is a reflecting wall at $x=L$ (we first expose the method in $d=1$). In the simulations, only the first monomer is affected by the presence of the target and  of the reflecting walls. The initial value of a mode $a_i$ ($i\ge2$) is taken from a normal distribution of variance $1/\lambda_i$. The initial value of $a_1$ is chosen such that the position of the first monomer is $x_1^0$. The initial position of the monomers is then obtained by applying Eq. (\ref{DefinitionModes}). The, the position of the monomers $x_i,i\ge2$ evolves at each time step with: 
\begin{equation}
	x_i(t_{k+1})=	x_i(t_{k}) -\left[\sum_{j=1}^N M_{ij} x_j(t_k)\right] (\Delta t) +  \sqrt{2\ (\Delta t)}\ u_{i,k}
\end{equation}
in which $u_{i,k}$ is a random number taken from a gaussian distribution with variance 1. If $x_1(t_k)$ is far from the boundaries, it also evolves according to this equation: 
\begin{equation}
	x_1(t_{k+1})=	x_1(t_{k}) +F_1 \Delta t +  \sqrt{2\ \Delta t}\ u_{1,k} \ ; \ F_1=-\sum_{j=1}^N M_{1j} x_j(t_k)
\end{equation}
If the position $x_1$ is close from the reflecting wall, then the last equation has to be modified in order to take into account the fact that the particle has a decreased probability to approach the wall (and zero probability to cross it). Therefore, if $\vert L-x_1\vert <d$, we follow the procedure introduced in Ref. \cite{Peters2002}:
\begin{equation}
x_1(t_{k+1})=x_1(t_k)-f_1^{\text{refl}}\left(\frac{L-x_1}{\sqrt{\Delta t}}\right)\sqrt{\Delta t}+u_{1,k}\ f_2^{\text{refl}}\left(\frac{L-x_1}{\sqrt{\Delta t}}\right)\Delta t + F_1\Delta t \label{Eq908}
\end{equation}
In this equation, $u_{1,k}$ is a random number that takes the values $\pm1$ with equal probability, and the positive functions $f_1^{\text{refl}}$ and $f_2^{\text{refl}}$ are the functions $f_1$ and $f_2$ of the equation (18) in the reference \cite{Peters2002}. Similarly, if $x_1$ is close from the absorbing boundary (\textit{i.e.} $x_1<d$), one calculates $P_{\text{abs}}=1-\text{erf}(x_1/(2\sqrt{\Delta t}))$ the probability of being absorbed between $t$ and $t+\Delta t$. One then generates a random number between 0 and 1 to decide whether or not the target is reached during the time step, in which case the simulation stops. If the absorbing wall is not reached, then $x_1$ evolves according to:
\begin{equation}
x_1(t_k+1)=x_1(t_k)+f_1^{\text{abs}}\left(\frac{x_1}{\sqrt{\Delta t}}\right)\sqrt{\Delta t}+ u_{1,k} \ f_2^{\text{abs}}\left(\frac{x_1}{\sqrt{\Delta t}}\right)\Delta t + F_1\Delta t \label{Eq67}
\end{equation}
where the random number $u_{1,k}$ takes again the values $\pm 1$ with equal probability. The positive functions $f_1^{\text{abs}}$ and $f_2^{\text{abs}}$ are the functions $f_1$ and $f_2$ of the equation (16) in the reference \cite{Peters2002}.
At the end of the simulation, the positions of the monomers are recorded, thereby giving an access to the splitting probability. The parameters used in the simulations are: $d=L/10$, $\Delta t=5\times 10^{-4}$, $L=158$ (the volume is $V=2L=316$), $N=10$. 

In 3 dimensions, we apply the same algorithm except that the equations (\ref{Eq908},\ref{Eq67}) are applied to the radial component of the position of the first monomer. The target is  a sphere of radius $a=1.7$ at the center of the volume (sphere of radius $R=28.4$). The time step for these simulations is also $\Delta t=5\times 10^{-4}$, and the distance $d$ is $2.85$. In all the simulations, the random numbers are generated with the ``ran2'' algorithm described in the \textit{Numerical Recipes in C} \cite{PressBook}.

\bibliographystyle{naturemag.bst}

%\bibliography{BiblioNonmarkovianPolymer}
%\bibliography{BiblioPostDoc}
%\bibliography{/Users/thomasguerin/Documents/Recherche/Biblio/BiblioPostDoc/BiblioPostDoc}

\end{document}